\documentclass[prd,aps,twocolumn,a4paper,floatfix]{revtex4-2}

\usepackage{graphicx,psfrag,mathrsfs}
\usepackage{slashed}
\usepackage{mathrsfs}
\usepackage{amsmath,amsfonts,amssymb,amsthm,array}
\usepackage{hyperref}
\hypersetup{breaklinks=true}
\usepackage{url}
\usepackage{comment,cancel}
\usepackage{accents}
\usepackage{ulem}
\usepackage{xcolor}

\makeatletter
\newcommand{\sbullet}{%
  \hbox{\fontfamily{lmr}\fontsize{.5\dimexpr(\f@size
pt)}{0}\selectfont\textbullet}}
\DeclareRobustCommand{\mathbullet}{\accentset{\sbullet}}
\makeatother

\def\p{\partial}  \def\ul{\underline}
\def\non{\nonumber} \def\mn{\mathring{\nabla}}
\def\mbn{\mathbullet{\nabla}}
\def\Gmmb{\Gamma[\mathring{\nabla},\mathbullet{\nabla}]}
\def\Gb{\mathbullet{\Gamma}} \def\sD{\slashed{D}} \def\sg{\slashed{g}}
\def\psib{\ul{\psi}} \def\sigmab{\ul{\sigma}} \def\cmr{\mathcal{C}^R_-}

\newtheorem{thm}{Theorem}

\newtheorem{remark}{Remark}

\begin{document}

\title{Peeling in Generalized Harmonic Gauge}

\author{Miguel Duarte$^{1,2}$}
\author{Justin Feng$^2$}
\author{Edgar Gasper\'in$^{2,3}$}
\author{David Hilditch$^2$}

\affiliation{$^1$CAMGSD, Departamento de Matem\'atica, Instituto
  Superior T\'ecnico IST, Universidade de Lisboa UL, Avenida Rovisco
  Pais 1, 1049 Lisboa, Portugal, $^2$CENTRA, Departamento de F\'isica,
  Instituto Superior T\'ecnico IST, Universidade de Lisboa UL, Avenida
  Rovisco Pais 1, 1049 Lisboa, Portugal, \\$^3$ Institut  de
  Math\'ematiques  de  Bourgogne  (IMB),  UMR  5584, CNRS,Universit\'e
  de  Bourgogne  Franche-Comt\'e, F-21000  Dijon,  France}

\begin{abstract}
It is shown that a large class of systems of non-linear wave
equations, based on the {\it good-bad-ugly} model, admit formal
solutions with polyhomogeneous expansions near null infinity. A
particular set of variables is introduced which allows us to write the
Einstein field equations in generalized harmonic gauge as a {\it
  good-bad-ugly} system and the functional form of the first few
orders in such an expansion is found by applying the aforementioned
result. Exploiting these formal expansions of the metric components,
the peeling property of the Weyl tensor is revisited. The question
addressed is whether or not the use of generalized harmonic gauge, by
itself, causes a violation of peeling. Working in harmonic gauge, it
is found that log-terms that prevent the Weyl tensor from peeling do
appear. The impact of gauge source functions and constraint additions
on the peeling property is then considered. Finally, the special
interplay between gauge and constraint addition, as well as its
influence on the asymptotic system and the decay of each of the metric
components, is exploited to find a particular gauge which suppresses
this specific type of log-term to arbitrarily high order.
\end{abstract}

\maketitle

\section{Introduction}

The concept of null infinity plays a very important role in general
relativity (GR) and
astrophysics~\cite{Pen63,NewPen62,BonBurMet62,Sac62,Sac62a}. From the
point of view of both numerical and mathematical relativity, it is
necessary in finding solutions to long-standing problems such as the
weak cosmic censorship conjecture, global stability analysis of
spacetimes and their (non) peeling properties. We will come back to
the latter later in this introduction, as it is the main topic that
this work aims to address.  From an astrophysics point of view, null
infinity is also a vital concept, since gravitational radiation is
non-localizable and is hence only well-defined at null
infinity. Waveforms computed in numerical relativity are however
evaluated at large radius and extrapolated to infinity. There are
various approaches to avoid this, and instead to include null infinity
in the computational domain. For instance, the work of
H\"ubner~\cite{Hub99,Hub01} and Frauendiener~\cite{DouFra16} makes use
of the conformal Einstein field equations (CEFE) introduced by
Friedrich~\cite{Fri81,Fri81a}, in turn building upon Penrose's idea of
bringing null infinity to a finite coordinate distance by means of a
conformal compactification~\cite{Pen63}. Although the CEFE provide a
geometric approach to the problem of the inclusion of null infinity,
the standard methods of numerical relativity that have proven to work
well for the strong field region of spacetimes of physical interest
cannot be {\it trivially} lifted over. In particular, this approach
has not yet been used for compact binary evolutions.

Chief alternative approaches to the problem are
Cauchy-characteristic-matching~\cite{Win05}, and the use of initial
data on a hyperboloidal slice. Such slices are not Cauchy
hypersurfaces, as their domain of dependence does not contain the
entire spacetime, rather they are spacelike hypersurfaces which
intersect null infinity. At the PDE level, this presents technical
difficulties, the main issue being the appearance of formally singular
terms. Numerical treatments of this in spherical symmetry
are~\cite{Zen07,VanHusHil14,VanHus14,Van15}. Further hyperboloidal
formulations of GR (with successful numerics) can be found
in~\cite{BucPfeBar09,BarSarBuc11,MonRin08,RinMon13}.

It is thus desirable to find formulations that mollify the formally
singular terms. A recent proposal is the dual frame
approach~\cite{Hil15}. This strategy consists of decoupling the
coordinates from the tensor basis and carefully choosing each, thus
allowing one to write the Einstein field equations (EFE) in
generalized harmonic gauge (GHG) and then solve the resulting
equations in hyperboloidal
coordinates~\cite{HilHarBug16,GasHil18,GasGauHil19,GauVanHil21}.
Important ingredients in this method are the coordinate lightspeed
condition and the weak null condition~\cite{GasHil18,LinRod03}. The
former is the requirement that derivatives of the radial coordinate
lightspeeds have a certain fall-off near null infinity, while the
latter is expected to be a sufficient condition on the non-linearities
of a quasilinear wave equation for establishing small data global
existence. Although it has not been shown in full generality that the
weak null condition implies small data global existence, recent work
by Keir~\cite{Kei17} proved that if a system of quasilinear wave
equations satisfies the \textit{hierarchical weak null condition},
then small data global existence is guaranteed. In~\cite{Kei19}, the
conditions required on the nonlinearities were relaxed even further.

It has been shown in~\cite{HilHarBug16} that the dual foliation
formulation of GR in GHG, together with hyperboloidal coordinates, can
be used to avoid the worst type of formally singular terms. In
harmonic gauge, however, even the simplest choices of initial data
give rise to terms with decay near null infinity of the
type~$O(R^{-1}\log R)$, with~$R$ a suitably defined radial
coordinate. Such terms are problematic in numerical evolutions, which
eventually {\it see} logarithms as an obstruction to rapid
convergence. In~\cite{GasGauHil19}, the authors used a toy model
composed of wave equations with non-linearities of the same kind as
those present in the EFE to show that these logarithmically divergent
terms can be explicitly regularized by a non-linear change of
variables. This toy model is called the \textit{good-bad-ugly} model
as it splits the evolved fields into three categories according to
their fall-off near null infinity, and it is known to satisfy the weak
null condition.

Exploiting conformal transformations and under the assumption of the
smoothness of null infinity, it can be shown, under very general
conditions, that the components of the Weyl tensor of spacetimes
associated with an isolated source fall off with certain negative
integer powers of~$r$, an affine parameter along outgoing null
geodesics ~\cite{NewPen62}. This property, known as \textit{peeling},
implies in particular that the far field of any source of
gravitational waves behaves locally as a plane wave. The question of
whether physically relevant spacetimes satisfy the peeling property
was a subject of extensive debate. Whereas solutions of interest have
been shown to allow for a smooth null infinity, namely using
hyperboloidal initial data~\cite{Fri86,Fri83,AndChrFri92}, and
asymptotically flat initial data~\cite{ChrDel02,Cor07}, it is now
clear that several physically motivated constructions violate peeling,
and hence cannot have smooth null
infinity~\cite{Keh21,Keh21_a}. Rigorous evidence (from the PDE
perspective) of logarithmic terms in the asymptotics of solutions to
the EFE in harmonic gauge can be found in \cite{Lin17} and in the
proof of the stability of the Minkowski spacetime (with
polyhomogeneous initial data and developments) of~\cite{HinVas20}.

In~\cite{DuaFenGasHil21}, a heuristic method to find asymptotic
solutions to the good-bad-ugly system was laid out. The
non-linearities in this model are known to mimic those present in the
EFE. Formal expansions are derived in which terms proportional to the
logarithm of the radial coordinate appear at every order in the bad
field, from the second order onward in the ugly field but never in the
good field. The model was then generalized to wave operators built
from an asymptotically flat metric and it was shown that it admits
polyhomogeneous asymptotic solutions. Finally the authors define
stratified null forms, a generalization of standard null forms, which
capture the behavior of different types of fields, and demonstrate
that the addition of such terms to the original system bears no
qualitative influence on the type of asymptotic solutions found.

In this work we build upon~\cite{DuaFenGasHil21} by pursuing the same
strategy to find polyhomogeneous expansions of the EFE near null
infinity. The question addressed is whether or not, even within a
favorable class of initial data, the peeling property is manifest in
harmonic or generalized harmonic gauge. In section~\ref{setup} we
review the formalism we use, which is very similar to that used
in~\cite{DuaFenGasHil21}. We furthermore write the derivatives of
metric components in terms of components of the transition tensor
between two covariant derivatives. In section~\ref{GBUwSNF} we state
the theorem shown in~\cite{DuaFenGasHil21} and generalize its proof to
include GR in GHG. In section~\ref{EFE} we derive the EFE as a system
of 10 coupled non-linear wave equations for our variables of
choice. Section~\ref{AsCartHarm} is dedicated to showing that the EFE,
in Cartesian harmonic gauge and with a particular constraint addition,
constitutes a good-bad-ugly system. Building on previous results, we
then integrate the equations for the first few orders to find the
functional form of the leading terms in the polyhomogeneous
expansion. Informally, we integrate the equations to find the powers
of~$\log R$ that can be present in each term with general initial
data. In section~\ref{NoPeeling} we show that there are choices of
initial data which prevent the Weyl scalars from peeling by
introducing a~$\log R/R^3$ term in~$\Psi_2$. In
section~\ref{recoverpeeling} we explore the interplay between gauge,
constraint addition and decay. Inspired by the general strategy
of~\cite{LinMatRin07}, we show that there is a particular choice of
gauge source function which recovers the peeling property by
preventing any powers~$\log R$ from appearing in the first orders of
the expansion. In section~\ref{logsback}, we show that by applying a
change of coordinates to this good gauge choice, the obstruction to
peeling that we saw earlier reappears. We conclude in
section~\ref{conclusions}.

\section{Geometric Set up}\label{setup}

\paragraph*{Representation of the metric:}
Abstract tensor indices are represented by Latin letters and
coordinate indices are represented by Greek letters from the beginning
of the alphabet. We assume a Lorentzian metric~$g_{ab}$ with
Levi-Civita connection~$\nabla$; indices are raised and lowered with
the spacetime metric~$g_{ab}$ and it's inverse. We introduce an
asymptotically Cartesian coordinate system
$X^{\ul{\alpha}}=(T,X^{\ul{i}})$, with the respective vector and
covector bases $\p_{\ul{\alpha}}$ and~$dX^{\ul{\alpha}}$. The sense in
which $X^{\ul{\alpha}}=(T,X^{\ul{i}})$ is asymptotically Cartesian
will be clarified later. It is useful to define the vector field
$T^a:=\p_T^a$, with the associated directional derivative denoted
$\nabla_T=T^a\nabla_a$. Directional derivatives along other vector
fields will be defined similarly. The flat covariant derivative
associated to $X^{\ul{\alpha}}$ is denoted $\mn$ and has the defining
property that $\mn_a\p_{\ul{\alpha}}^b=0$. We also introduce shell
coordinates~$X^{\ul{\alpha}'}=(T',X^{\ul{i}'})=(T,R,\theta^A)$, with
radial coordinate given by
$R^2=(X^{\ul{1}})^2+(X^{\ul{2}})^2+(X^{\ul{3}})^2$ and the respective
vector and covector bases $\p_{\ul{\alpha}'}$ and~$dX^{\ul{\alpha}'}$.
The flat covariant derivative associated to the shell coordinates has
the defining property that $\mbn_b\p_{\ul{\alpha}'}^a=0$. The flat
covariant derivatives $\mn$ and $\mbn$ reduce to partial derivatives
when working in the associated coordinate bases. These definitions are
based on the notation used in xAct~\cite{xAct_web_aastex} to represent
partial derivatives using abstract index notation. Given the two flat
covariant derivatives $\mn$ and $\mbn$, we define the Christoffel
transition tensor
\begin{align}
\Gmmb_{a}{}^{b}{}_{c}v^c = \mn_av^b - \mbn_av^b\,,
\end{align}
$v^a$ being an arbitrary vector field. Similarly,
\begin{align}
\Gamma[\nabla,\mathbullet{\nabla}]_{a}{}^{b}{}_{c}v^c = \nabla_av^b -
\mbn_av^b\,,
\end{align}
which will be represented with the shorthand~$\Gb_{a}{}^{b}{}_{c} :=
\Gamma[\nabla,\mathbullet{\nabla}]_{a}{}^{b}{}_{c}$. It follows that
\begin{align}\label{eq:CartesianAndShellConnectionRelation}
  \Gamma[\nabla,\mathring{\nabla}]_b{}^a{}_c
  =\Gamma[\nabla,\mathbullet{\nabla}]_b{}^a{}_c -
  \Gamma[\mathring{\nabla},\mathbullet{\nabla}]_b{}^a{}_c.
\end{align}
For clarity, we show these coordinates and their covariant derivatives
in the following table:
\begin{center}
\begin{tabular}{  m{7em} | c  } 
  coordinates & cov. der.  \\ 
  \hline
   $X^{\ul{\alpha}}=(T,X^{\ul{i}})$ & $\mn$ \\ 
  \hline
  $X^{\ul{\alpha}'}=(T',X^{\ul{i}'})$ & $\mbn$  \\ 
\end{tabular}
\end{center}
We define outgoing and incoming null vectors on the spacetime
($\psi^a$ and~$\ul{\psi}^a$, respectively) in the Shell coordinate
basis by
\begin{align}\label{eq:psiinshellchart}
  \psi^a&=\p_T^a+\mathcal{C}_+^R\p_R^a\,,\nonumber\\
  \ul{\psi}^a&=\p_T^a+\mathcal{C}_-^R\p_R^a\,,
\end{align}
where~$\mathcal{C}_+^R$ and~$\mathcal{C}_-^R$ are radial coordinate
light-speeds, the values for which are determined by the condition
that~$\psi^a$ and~$\ul{\psi}^a$ are null with respect to~$g_{ab}$. It
is useful to define the following covectors,
\begin{align}\label{eq:xitoeta}
\sigma_a&=e^{-\varphi}\psi_a\,,\quad
\ul{\sigma}_a=e^{-\varphi}\ul{\psi}_a\,,
\end{align}
with~$\varphi$ specified by the
condition~$\sigma_a\p_R^a=-\ul{\sigma}_a\p_R^a=1$. The covectors may
be decomposed in the following way:
\begin{align}\label{eq:xiinshellchart}
  \sigma_a&=-\mathcal{C}_+^R\nabla_aT+\nabla_aR +
  \mathcal{C}^+_A\nabla_a\theta^A\,,\nonumber\\
  \ul{\sigma}_a&=\mathcal{C}_-^R\nabla_aT-\nabla_aR
  + \mathcal{C}^-_A\nabla_a\theta^A\,.
\end{align}
It is straightforward to show that the null vectors satisfy the
relations
\begin{align}
  &\sigma_a\psi^a=\ul{\sigma}_a\ul{\psi}^a=0\,,\non\\
  &\sigma_a\ul{\psi}^a=\ul{\sigma}_a\psi^a=-\tau\,,
\end{align}
with~$\tau:=\mathcal{C}_+^R-\mathcal{C}_-^R$.

We perform the following decomposition of the inverse spacetime
metric,
\begin{align}\label{eq:metricrepresentation}
g^{ab}=-2\tau^{-1}e^{-\varphi}\,\psi^{(a}\ul{\psi}^{b)}+\slashed{g}^{ab}
\,,
\end{align}
with the normalization of first term chosen so
that~$\slashed{g}^{ab}\sigma_b=\slashed{g}^{ab}\ul{\sigma}_b=0$. We
emphasize that while~$\slashed{g}_a{}^b$ serves as a projection
operator orthogonal to~$\sigma_a$ and~$\ul{\sigma}_b$,
$\slashed{g}^{ab}$ is not the inverse induced metric on level sets
of~$T$ and~$R$, as~$\nabla_aT$ or~$\nabla_aR$ are not in the kernel of
$\slashed{g}^{ab}$. We obtain the following decomposition for the
metric,
\begin{align}
  g_{ab}=-2\tau^{-1}e^{\varphi}\,\sigma_{(a}\ul{\sigma}_{b)}+\slashed{g}_{ab}\,.
\end{align}
Let~$\mathcal{S}$ be a Cauchy surface defined by a constant value of
the coordinate~$T$. It will be convenient to make a conformal
rescaling of this induced metric on~$\mathcal{S}$. We define
\begin{equation}\label{eq:qconformaltransformation}
  q_{ab}= e^{-\epsilon}R^{-2}\slashed{g}_{ab},
  \qquad (q^{-1})^{ab} = e^{\epsilon}R^{2}\slashed{g}^{ab}
\end{equation}
with
\begin{equation}
\epsilon = (\ln|\slashed{g}|-\ln|\slashed{\mathbullet{g}}| )/2
\end{equation}
where~$ |\slashed{\mathbullet{g}}|$ denotes the determinant of the
metric of~$\mathbb{S}^2$ of radius~$R$ embedded in Minkowski spacetime
in Shell coordinates. For future reference we also
define~$\mathbullet{\epsilon} \equiv \tfrac{1}{2}\ln
|\slashed{\mathbullet{g}}|$. Notice that, as a consequence of the
latter, the determinant of~$q$ is that of the metric of the
unit~$\mathbb{S}^2$ in Shell coordinates, so it is a fixed function of
the coordinates~$\theta^A$. Although this construction is general for
any coordinatization of~$\mathbb{S}^2$, for conciseness, in the rest
of the paper, we use standard spherical polar
coordinates~$\theta^1=\theta$ and~$\theta^2=\phi$ so that the line
element on~$\mathbb{S}^2$ reads~$d\Omega^2= d\theta^2+\sin^2\theta
d\phi^2$. We have used the notation~$(q^{-1})^{ab} $ to emphasize that
the metric~$\slashed{g}$ and not~$q$ is used to raise and lower
indices on tensors on~$\mathcal{S}$. For instance~$q^{ab} =
\slashed{g}^{ac}\slashed{g}^{bd}q_{cd} \neq (q^{-1})^{ab}$. Finally,
it is worth introducing a special parameterization of the angular part
of $(q^{-1})^{ab}$ in Shell coordinates, in order to find wave
equations for each of its two independent metric functions. These
independent components encode the two degrees of freedom of
gravitational waves usually denoted in the linear theory as~$h_+$
and~$h_\times$.
\begin{align}
	(q^{-1})^{AB} = \begin{bmatrix}
e^{-h_+}\cosh h_\times & \frac{\sinh h_\times}{\sin \theta} \\
\frac{\sinh h_\times}{\sin \theta}
& \frac{e^{h_+}\cosh h_\times}{\sin \theta ^2}
\end{bmatrix}\,.
\end{align}
Thus the ten independent variables we use to represent the metric are,
\begin{align}\label{eq:BasicMetricVariables}
 \varphi\,, \quad\mathcal{C}_\pm^R\,, \quad \mathcal{C}^\pm_A\,, \quad
  \epsilon\,, \quad h_+\,, \quad h_\times\,.
\end{align}
Tensors~$T'_{cd}$ projected with~$\slashed{g}_a{}^b$
are~$\slashed{T}_{ab} \equiv
\slashed{g}_a{}^c\slashed{g}_{b}{}^dT'_{cd}$. The projected covariant
derivative is denoted~$\sD$, and is given explicitly below for vectors
satisfying~$v^a=\slashed{g}_{b}{}^{a}v^b$,
\begin{align}
\sD_b v^a := \slashed{g}_{c}{}^{a}\slashed{g}_{b}{}^{d}\nabla_d v^c\,.
\end{align}
The generalization for higher rank tensors is straightforward. We
define the projected covariant derivative~$\mathring{\sD}$ similarly,
with~$\nabla_d$ replaced with~$\mn_d$.

A brief remark on terminology---since we deal with polyhomogeneous
expansions, which include terms of the form~$R^{-n}(\log R)^m$, it is
appropriate to clarify what is meant by the term `order' in this
article. Throughout, `order~$n$' will refer to terms proportional
to~$R^{-n}$.

\paragraph*{Decomposition of the connection:}
Using the Koszul formula,
\begin{align}
	\Gb_a{}^b{}_c = \frac{1}{2}\sg^{bd}(\mbn_a \sg_{db}+\mbn_b
        \sg_{ad}-\mbn_d \sg_{ab})\,,
\end{align}
we can find expressions for each component of~$\Gb_a{}^b{}_c$. Here we
present a rearranged version of these relations, writing derivatives
of metric functions in terms of components of the connection and not
the other way around. The reason for us to do this is that writing the
EFE in terms of~$\Gb_a{}^b{}_c$ makes them significantly more
readable. We find that,
\begin{align}\label{ConnectionComps}
  &\nabla_a \mathcal{C}_+^R =
  -\Gb_a{}^\sigma{}_\psi\quad\,,\quad\nabla_a \mathcal{C}_-^R =
  \Gb_a{}^{\sigmab}{}_{\psib}\,,\non\\ &\nabla_a \mathcal{C}^+_A =
  2\sg_{bA}\Gb_a{}^{(b
    \sigma)}-\frac{\mathcal{C}^+_A+\mathcal{C}^-_A}{\tau}
  \Gb_a{}^\sigma{}_\psi\,,\non\\ &\nabla_a
  \mathcal{C}^-_A = 2\sg_{bA}\Gb_a{}^{(b
    \sigmab)}-\frac{\mathcal{C}^+_A+\mathcal{C}^-_A}{\tau}
  \Gb_a{}^{\sigmab}{}_{\psib}\non\,,\\
  &\nabla_a
  \varphi = \Gb_a{}^b{}_c(\delta_b^c-\sg_b{}^c)\,,\nabla_a \sg^{AB} =
  -2\sg_b{}^A\sg_c{}^B\Gb^{(bc)}{}_a\,,\non\\ &\nabla_a (\epsilon +
  \mathbullet{\epsilon}) =\sg_c{}^b\Gb_b{}^c{}_a\,.
\end{align}
Note that these covariant
derivatives are interchangeable with~$\mbn$ and~$\mn$ even when they
act upon $\mathcal{C}^\pm_A$, as~$A$ is not a tensorial index, but a
label to designate one of the angular coordinates, $\theta^1=\theta$
or $\theta^2=\phi$. When writing the asymptotic system near null
infinity, we will have to use the version of these relations that
gives the components of the transition tensor~$\Gb_a{}^b{}_c$ in terms
of first derivatives of metric functions. In order not to overload the
reader with long expressions, and because they contain basically the
same information as~\eqref{ConnectionComps}, we do not present those
relations here.

\paragraph*{D'Alembert operators:}
Throughout this work we will use three different wave operators, each
associated to a different covariant derivative. Thus we
define~$\square$ as,
\begin{align}
  \square \phi = g^{ab}\nabla_a\nabla_b\phi\,,
\end{align}
and we define~$\mathring{\square}$ and $\mathbullet{\square}$
analogously, associated with~$\mn$ and~$\mbn$, respectively. Note that
when $\mathring{\square}$ acts on a scalar function, one can
straightforwardly change to~$\mathbullet{\square}$ through,
\begin{align}
  \mathring{\square} \phi = \mathbullet{\square} \phi -
  g^{bc}\Gamma[\mathring{\nabla},
    \mathbullet{\nabla}]_b{}^{a}{}_c\nabla_a\phi\,.
\end{align}
Moreover, in~\cite{DuaFenGasHil21} it was shown that the Cartesian wave
operator can be expanded as,
\begin{align}\label{lhs2}
  \mathcal{C}_+^R\mathring{\square} \phi = &-
  2e^{-\varphi}\nabla_{\psi}\nabla_{T} \phi
  +\nabla_T\phi(\slashed{g}^{ab}\Gmmb_{a}{}^{\sigma}{}_{b} + X_T)\non\\
  &+\nabla_{\psi}\phi X_\psi
  -\frac{2e^{-\varphi}\mathcal{C}_-^R}{\tau}\nabla_{\psi}^2\phi
  + \mathcal{C}_+^R\cancel{\Delta}\phi\,,
\end{align}
where~$X_T$ and~$X_\psi$ are,
\begin{align}
  \tau X_T:= &\mathcal{C}_A\sD^A\mathcal{C}_+^R
  - \tau\sD^A\mathcal{C}_A^+
  - \frac{2e^{-\varphi}\mathcal{C}_-^R}{\mathcal{C}_+^R}
  \nabla_\psi\mathcal{C}_+^R\,,\\\non
  \tau X_\psi:=& \frac{\mathcal{C}_A}{\tau}\left(\mathcal{C}_-^R
  \sD^A\mathcal{C}_+^R-\mathcal{C}_+^R
  \sD^A\mathcal{C}_-^R\right)
  - \mathcal{C}_-^R\sD^A\mathcal{C}_A^+\non\\
  &- \mathcal{C}_+^R\sD^A\mathcal{C}_A^-
  + \mathcal{C}_-^R\slashed{g}^{ab}\Gmmb_{a}{}^{\sigma}{}_{b}\non\\
  &+ \mathcal{C}_+^R\slashed{g}^{ab}\Gmmb_{a}{}^{\ul{\sigma}}{}_{b}
  + \frac{2e^{-\varphi}\mathcal{C}_-^R}{\mathcal{C}_+^R}
  \nabla_\psi\mathcal{C}_+^R \,,\non
\end{align}
and~$\mathcal{C}_A:=\mathcal{C}_A^++\mathcal{C}_A^-$. Most
importantly, it was shown that, to leading order, this operator
behaves as,
\begin{align}\label{lhsLeading}
  \mathring{\square} \phi \simeq -2\nabla_{\psi}\nabla_{T} \phi
  -\frac{2}{R}\nabla_T\phi\,,
\end{align}

\section{GBU system with stratified null forms}\label{GBUwSNF}

\textit{Stratified null forms} are defined as terms that involve
products of up to one derivative of the evolved fields and fall-off
faster than~$R^{-2}$ close to null
infinity. Henceforth~$\mathcal{N}_\phi$, where~$\phi$ is the evolved
field that~$\mathcal{N}_\phi$ is associated to, will denominate an
arbitrary linear combination of stratified null forms.

\paragraph*{The \textit{good-bad-ugly} system:} We introduce
the following model,
\begin{align}
  & \mathring{\square} g = \mathcal{N}_g\,,\non\\
  &\mathring{\square} b = (\nabla_T g)^2 + \mathcal{N}_b\,,\non\\
  &\mathring{\square} u = \tfrac{2}{R}\nabla_T u + \mathcal{N}_u\,,
  \label{gbu}
\end{align}
where~$g$, $b$ and~$u$ stand for~{\it good, bad} and {\it ugly}
fields, respectively. The leading order of the particular case where
the metric~$g^{ab}$ is the Minkowski metric and~$\mathcal{N}_\phi=0$
was studied in detail in~\cite{GasGauHil19}. The main reason for the
interest in this particular model lies in the fact that its
non-linearities are known to mimic those present in the EFE. Previous
work~\cite{DuaFenGasHil21} has shown that this system admits
polyhomogeneous expansions near null infinity. In this work we want to
build directly upon that result, by applying it to GR and finding the
functional form of terms beyond first order. Then we aim to shed light
on a very special interplay between the choice of gauge and constraint
addition by showing that peeling is violated for certain choices and
satisfied in others. Additionally, this will allow us to know where
terms with factors of $\log R$ may appear and thus builds towards a
full-blown regularization of the EFE in GHG in hyperboloidal slices
within the Dual Foliation framework. Finally, we expect that this
method is also applicable to the Maxwell and Yang-Mills equations, as
these can also be written as non-linear wave equations in Lorenz
gauge.

\subsection{Assumptions}\label{assumptions}

To proceed, it is appropriate to outline assumptions on the evolved
fields in~\eqref{gbu} that allow us to formally equate terms of the same
order and identify a hierarchy of equations which are satisfied by the
fields~$g$, $b$ and~$u$ order-by-order close to null infinity. We also
discuss corresponding assumptions for the metric functions.

\paragraph*{Evolved fields:}
Consider a null tetrad ~$\{\psi,\ul{\psi},X_1,X_2\}$, with~$X_1$
and~$X_2$ orthogonal to~$\psi^a$ and~$\ul{\psi}^a$ and normalized so
that $g_{ab}X_A^a\psi^b=g_{ab}X^a_A\ul{\psi}^b=0$
and~$g_{ab}X_A^aX_B^b=\delta_{AB}$, with $A=1,2$. Now
let~$\omega_{g,b}$ represent a good or a bad field or any of its first
derivatives. Following the insight of~\cite{Kei17} we assume the
following behavior at null infinity,
\begin{align}\label{weaknull1}
\omega_{g,b} = o^+(R^{-n})\Rightarrow
\begin{cases}
  \nabla_{\psi}\omega_{g,b} = o^+(R^{-n-1})\\
  \nabla_{\ul{\psi}}\omega_{g,b} = o^+(R^{-n})\\
  \nabla_{X_A}\omega_{g,b} = o^+(R^{-n-1})
\end{cases}
\,.
\end{align}
The notation~$f=o^+(h)$ refers to the condition
\begin{align}\label{deflittleo}
  \exists \epsilon>0 :
  \lim_{R\rightarrow\infty} \frac{f}{hR^{-\epsilon}}=0\,,
\end{align}
which can be informally stated as the condition that \textit{$f$
  falls-off faster than~$h^{1+\epsilon}$ as~$R$ goes to infinity},
which is a faster falloff than~$f=o(h)$; more
precisely,~$o^+(h)=o(hR^{-\epsilon})$. As discussed
in~\cite{DuaFenGasHil21} this condition will be needed to ensure that
the error terms remain small when integrated. We employ the
condition~$o^+(R^{-n})$ instead of~$O(R^{-n-1})$ since it was shown
previously (for instance~\cite{GasGauHil19}) that the
system~\eqref{gbu} admits asymptotic solutions proportional
to~$R^{-1}\log R$, and the~$O$ notation naively excludes such
solutions. Notice that for fields satisfying a system of the
form~\eqref{gbu}, certain derivatives improve the fall off of the
argument, but others derivatives do not; this motivates the
terminology employed in the previous literature, namely that
directional derivatives corresponding to~$\psi^a$ and~$X_A$ (those
tangent to outgoing null-cones) are termed good derivatives, and
directional derivatives corresponding to~$\ul{\psi}^a$ (transverse to
outgoing null-cones) are termed bad derivatives.

Now let~$\omega_u$ represent the field~$u$ or its first derivative. As
discussed in~\cite{GasGauHil19}, if~$\mathring{\square}$ is
constructed from the Minkowski metric, the derivatives of the~$u$
fields have different asymptotic behavior; we assume the following,
\begin{align}\label{weaknull2}
  \omega_u = o^+(R^{-n})\Rightarrow
  \begin{cases}
    \nabla_{\psi}\omega_u = o^+(R^{-n-1})\\
    \nabla_{\ul{\psi}}\omega_u = o^+(R^{-n-1})\\
    \nabla_{X_A}\omega_u = o^+(R^{-n-1})
  \end{cases}
  \,.
\end{align}
We seek solutions which decay near null infinity, so we restrict
ourselves to initial data with that property, i.e.,
\begin{align}
	g=o^+(1),\quad b=o^+(1), \quad u=o^+(1)\,.
\end{align}
To allow for a nonzero ADM mass and linear momentum, the initial data
is chosen to decay at spacelike infinity (as opposed to null infinity)
in the following manner,
\begin{align}
&\phi\rvert_{\mathcal{S}} = \sum_{n=1}^{\infty}\frac{m_{\phi,n}}{R^n}\,,\non\\
  &\nabla_T \phi\rvert_{\mathcal{S}} = O_{\mathcal{S}}(R^{-2})\,,
  \label{assumptionTimeder}
\end{align}
where~$m_{\phi,n}$ are scalar functions that are independent of~$T$
and~$R$. Though this is not the most general choice which permits
nontrivial ADM mass and linear momentum, it is general enough for a
very large class of spacetimes of interest.

\paragraph*{Metric functions:}
We now turn to the metric functions. We require that the metric
functions are written as,
\begin{align}
  &\varphi = \gamma_1\quad,\quad\mathcal{C}_\pm^R =
  \pm1 + \gamma^\pm_2\quad,\quad\mathcal{C}_A^\pm =
  R\gamma_3^\pm\,, \non \\
  &\epsilon = \gamma_4\quad,\quad h_+ = \gamma_5
  \quad,\quad h_\times = \gamma_6\,,
\label{asympflatness}
\end{align}
where the~$\gamma=\gamma(g,b,u)$ are assumed to be analytic in a
neighborhood of null infinity~$\mathscr{I}^+$. Since we are interested
in metrics that asymptote to the Minkowski metric
near~$\mathscr{I}^+$, we require that
\begin{align}
	\gamma(g,b,u)|_{\mathscr{I}^+}=0\,,
\end{align}
or that the~$\gamma$ functions vanish as one approaches asymptotic
infinity. In fact, we will see that upon an appropriate choice of
gauge and constraint addition, in GR the~$\gamma$ functions turn out
to be themselves either good, bad or ugly. The metric variables in
this formulation were chosen for this purpose.

\subsection{Earlier work}

In~\cite{DuaFenGasHil21} the following theorem was shown,
\begin{thm} \label{T3} The \textit{good-bad-ugly} system defined
as~\eqref{gbu} where~$g_{ab}$ is an asymptotically flat metric, admits
a polyhomogeneous expansion near null infinity of the type,
\begin{align}\label{polyhomo}
  &g = \frac{G_{1,0}(\psi^*)}{R} +
  \sum_{n=2}^{\infty}\sum_{k=0}^{N_n^g} \frac{(\log R)^k
    G_{n,k}(\psi^*)}{R^n}\non\\ &b = \frac{B_{1,0}(\psi^*)+\log
    RB_{1,1}(\psi^*)}{R} + \sum_{n=2}^{\infty}\sum_{k=0}^{N_n^b}
  \frac{(\log R)^k B_{n,k}(\psi^*)}{R^n}\non\\ &u =
  \frac{m_{u,1}}{R}+\sum_{n=2}^{\infty}\sum_{k=0}^{N_n^u} \frac{(\log
    R)^k U_{n,k}(\psi^*)}{R^n} \,,
\end{align}
where~$\psi^*$ means that this scalar function does not vary along
integral curves of~$\psi$, $\Phi_{n,k}$ with~$\Phi\in \{G,B,U\}$ are
coefficients where~$n$ is their associated power of~$R^{-1}$ and~$k$
their associated power of~$\log R$, and with initial data
on~$\mathcal{S}$ of the type,
\begin{align}
  \begin{cases}
    g\rvert_{\mathcal{S}}=
    \sum_{n=1}^{\infty}\frac{m_{g,n}}{R^n}\\ b\rvert_{\mathcal{S}}=
    \sum_{n=1}^\infty \frac{m_{b,n}}{R^n}\\ u\rvert_{\mathcal{S}}=
    \sum_{n=1}^\infty \frac{m_{u,n}}{R^n}
  \end{cases}\,,
  \begin{cases}
    \nabla_Tg\rvert_{\mathcal{S}}=
    O_{\mathcal{S}}(R^{-2})\\ \nabla_Tb\rvert_{\mathcal{S}}=
    O_{\mathcal{S}}(R^{-2})\\ \nabla_Tu\rvert_{\mathcal{S}}=
    O_{\mathcal{S}}(R^{-2})
  \end{cases}\,,
\end{align}
where~$m_{\phi,n}$ are scalar functions that are independent of~$T$
and~$R$. This is valid outside a compact ball centered at~$R=0$.
\end{thm}

The assumptions on initial data could be relaxed if we wished only to
build formal solutions down to a finite order in~$R^{-1}$.

Throughout the article, all functions denoted by~$m_{\phi,n}$ for any
field~$\phi$ and any integer~$n$ will be assumed to be independent
of~$T$ and~$R$. We will show that in GR the metric
functions~\eqref{eq:BasicMetricVariables} can be separated into {\it
  good}, {\it bad} and {\it ugly} fields, as defined in~\eqref{gbu}.
Moreover, we will see that there is an interplay between gauge choice
and constraint addition that interferes with the asymptotic system and
gives rise to different combinations of these fields. However there
are three generalizations that we have to do to the above-mentioned
theorem in order to include GR with the gauge choices of
interest. First, our good-bad-ugly system should be allowed to include
arbitrary numbers of \textit{good}, \textit{bad} and \textit{ugly}
fields. In particular, in GR, there will be 10 independent metric
functions, each with its own nonlinear wave equation. Second, the
leading term on the RHS of the bad equation must be allowed to
be~$-1/2(\nabla_Tg_1)^2-1/2(\nabla_Tg_2)^2$, where~$g_1$ and~$g_2$ are
good fields. And third, GR forces us to extend our conception of an
ugly equation to include a slightly larger class of equations whose
asymptotic expansions behave as uglies to leading order, but differ in
the decay rate of terms beyond first order.

\subsection{Generalization of Theorem 1}
Let there be any number of evolved fields, each satisfying one of the
following 3 kinds of nonlinear wave equation,
\begin{align}\label{gbugeneral}
   &\mathring{\square} g = \mathcal{N}_g\,,\non\\ &\mathring{\square}
  b = \frac{\beta}{R^2} + \mathcal{N}_b\,,\non\\ &\mathring{\square} u
  = \frac{2p}{R}\nabla_T u + \mathcal{N}_u\,,
\end{align}
where $-2\beta=R^2(\nabla_Tg_1)^2+R^2(\nabla_Tg_2)^2$, $g_1$ and $g_2$
satisfy an equation of the first type, and $p$ is a natural number. To
prove a more general version of Theorem 1 we employ a method almost
exactly the same that considered in~\cite{DuaFenGasHil21}. For this
reason we leave out the common details and focus exclusively on the
differences.

\paragraph*{Motivation for induction hypothesis:} Asymptotic flatness,
together with the fact that the~$\gamma$ functions
(see~\eqref{asympflatness}) are analytic functions of the evolved
fields at null infinity, allows us to Taylor expand them
around~$g=b=u=0$, because the fields are assumed to have decay near
null infinity and find that~$\gamma=o^+(1)$ and,
\begin{align}\label{dersgamma}
  \omega_\gamma = o^+(R^{-n})\Rightarrow
  \begin{cases}
    \nabla_{\psi}\omega_\gamma =
    o^+(R^{-n-1})\\ \nabla_{X_A}\omega_\gamma = o^+(R^{-n-1})
  \end{cases}
  \,,
\end{align}
where~$\omega_\gamma$ is any~$\gamma$ function or any first derivative
of it. Let us rescale~$g$ and $b$ as,
\begin{align}\label{G1B1}
	\mathcal{G}_1=Rg\quad,\quad \mathcal{B}_1=Rb\,,
\end{align}
and focus on expansions of~$u$ which satisfy,
\begin{align}\label{umU2}
	u=\frac{m_{u,1}}{R} + \frac{\mathcal{U}_2}{R^2}\,,
\end{align}
with~$\mathcal{U}_2=o^+(R)$. Plugging~\eqref{G1B1} and~\eqref{umU2}
in~\eqref{gbugeneral} we get,
\begin{align}\label{g1b1u2eq}
  &\nabla_\psi\nabla_T\mathcal{G}_1 \simeq
  0\,,\non\\ &\nabla_\psi\nabla_T\mathcal{B}_1 \simeq
  -\frac{1}{R}\beta\,,\non\\ &
  \nabla_\psi(R^{p-1}\nabla_T\mathcal{U}_2) \simeq
  R^{p-2}\Omega^u_1\,,
\end{align}
where~$\Omega^u_1=o^+(R)$ cannot contain~$\mathcal{U}_2$ or any
derivatives thereof. For the first two equations we have,
\begin{align}\label{g1b1}
&\mathcal{G}_1 \simeq G_{1,0}(\psi^*)\non\\ &\mathcal{B}_1 \simeq
  B_{1,0}(\psi^*) + B_{1,1}(\psi^*)\log R\,.
\end{align}
The third equation requires closer attention. To leading order,
$\Omega^u_1$ is allowed to contain logs, but it is not allowed to
contain~$R^{-1}$, so we can write,
\begin{align}\label{omegaU1}
  \Omega^u_1 = \sum_{i=0}^{N_n^{\Omega^u}}
  (\log R)^i\Omega^u_{1,i}(\psi^*)\,.
\end{align}
In order to integrate the last equation in~\eqref{g1b1u2eq}
along~$\psi$, we have to integrate~$(\log R)^iR^{p-2}$, which gives
one of two results depending on~$p$,
\begin{align}
  \int \frac{(\log R)^i}{R^{-p+2}} dR = \begin{cases}
    \sum_{j=0}^i -\frac{(\log R)^j}{(-p+1)^{i-j+1}R^{-p+1}}
    \frac{i!}{j!}\,,\,p\neq1\\
    \frac{(\log R)^{i+1}}{i+1}\,,\,p=1
  \end{cases}.
\end{align}
The main difference between these two cases is that if~$p=1$, the
maximum power of~$\log R$ goes up by one, whereas if~$p\neq 1$ it does
not. Since we are allowing our null forms to have any finite power
of~$\log R$~\eqref{omegaU1}, this does not make a big difference for
the general case. However, more structure on these null forms will
allow us to find the power of~$\log R$ at each order. More on this
point will be said later. For now, for either case we can write,
\begin{align}
  \mathcal{U}_2 \simeq \sum_{i=0}^{N_2^{\Omega^u}}(\log R)^i
  U_{2,i}(\psi^*)+ \frac{1}{R^{p-1}}\int\dot{u}_2(\psi^*)dT\,,
\end{align}
where the second term can be incorporated in the first for~$p=1$,
while for~$p>1$ we choose solutions with~$\dot{u}_2(\psi^*)=0$, which
basically amounts to pushing the second term into a higher order term
in the expansion. This seems to suggest that~$g$, $b$ and~$u$ are
polyhomogeneous functions where each term can have up to~$N_n^\phi$
powers of~$\log R$, where~$N_n^\phi$ is a finite number which depends
on which field~$\phi$ it refers to and on the power of~$R^{-1}$ in the
expansion, $n$. Formally, we therefore conjecture,
\begin{align}\label{hyp}
  &g = \frac{G_{1,0}(\psi^*)}{R} +
  \sum_{n=2}^{\infty}\sum_{k=0}^{N_n^g} \frac{(\log R)^k
    G_{n,k}(\psi^*)}{R^n}\non\,,\\ &b = \frac{B_{1,0}(\psi^*)+\log
    RB_{1,1}(\psi^*)}{R} + \sum_{n=2}^{\infty}\sum_{k=0}^{N_n^b}
  \frac{(\log R)^k B_{n,k}(\psi^*)}{R^n}\non\,,\\ &u =
  \frac{m_{u,1}}{R}+\sum_{n=2}^{\infty}\sum_{k=0}^{N_n^u} \frac{(\log
    R)^k U_{n,k}(\psi^*)}{R^n}\,.
\end{align}
We proceed by induction. We already know that to first order in~$g$
and~$u$,~$\log R$ terms are not allowed, and the
conjecture~\eqref{hyp} incorporates this property by construction.
Truncating at~$n=1$, we have seen
\begin{align}
  &g = \frac{G_{1,0}(\psi^*)}{R}
  + \frac{\mathcal{G}_2}{R^2}\,,\non\\
  &b = \frac{B_{1,0}(\psi^*) + B_{1,1}(\psi^*)\log R}{R}
  + \frac{\mathcal{B}_2}{R^2}\,,\non\\
  &u = \frac{m_{u,1}}{R}+\frac{\mathcal{U}_2}{R^2}\,,
\end{align}
with~$\mathcal{G}_2=o^+(R)$, $\mathcal{B}_2=o^+(R)$
and~$\mathcal{U}_2=o^+(R)$, so in order to show~\eqref{hyp}, we have to
show that if we can write the evolved fields as,
\begin{align}\label{phin-1}
  &g = \frac{G_{1,0}(\psi^*)}{R} + \sum_{m=2}^{n-1}\sum_{k=0}^{N_{n-1}^g}
  \frac{(\log R)^k G_{m,k}(\psi^*)}{R^m}
  +\frac{\mathcal{G}_n}{R^n}\non\,,\\
  &b = \frac{B_1}{R}  + \sum_{m=2}^{n-1}\sum_{k=0}^{N_{n-1}^b} \frac{(\log R)^k
    B_{m,k}(\psi^*)}{R^m}+\frac{\mathcal{B}_n}{R^n}\non\,,\\
  &u =  \frac{m_{u,1}}{R}+\sum_{m=2}^{n-1}\sum_{k=0}^{N_{n-1}^u}
  \frac{(\log R)^k U_{m,k}(\psi^*)}{R^m}
  +\frac{\mathcal{U}_n}{R^n}\,,
\end{align}
where~$B_1=B_{1,0}(\psi^*)+\log
RB_{1,1}(\psi^*)$,~$\mathcal{G}_n=o^+(R)$, $\mathcal{B}_n=o^+(R)$
and~$\mathcal{U}_n=o^+(R)$, then we can also write them as,
\begin{align}\label{phin}
  &g = \frac{G_{1,0}(\psi^*)}{R} + \sum_{m=2}^{n}\sum_{k=0}^{N_n^g}
  \frac{(\log R)^k
    G_{m,k}(\psi^*)}{R^m}+\frac{\mathcal{G}_{n+1}}{R^{n+1}}\non\,,\\ &b
  = \frac{B_{1}}{R} + \sum_{m=2}^{n}\sum_{k=0}^{N_n^b} \frac{(\log
    R)^k
    B_{m,k}(\psi^*)}{R^m}+\frac{\mathcal{B}_{n+1}}{R^{n+1}}\non\,,\\
  &u = \frac{m_{u,1}}{R}+\sum_{m=2}^{n}\sum_{k=0}^{N_n^u}
  \frac{(\log R)^k U_{m,k}(\psi^*)}{R^m} +\frac{\mathcal{U}_{n+1}}{R^{n+1}}\,,
\end{align}
where~$\mathcal{G}_{n+1}=o^+(R)$, $\mathcal{B}_{n+1}=o^+(R)$
and~$\mathcal{U}_{n+1}=o^+(R)$. The cases of~$g$ and~$b$ go through in
exactly the same way as in~\cite{DuaFenGasHil21}, so we will focus our
attention on~$u$.

\paragraph*{Induction proof:} Assuming~\eqref{phin-1}, we take the ugly
equation in~\eqref{gbugeneral} and formally equate terms proportional
to $R^{-n-1}$. Putting all terms with~$\mathcal{U}_n$ on the LHS and
all the rest on the RHS we get,
\begin{align}\label{umot3}
\nabla_\psi(R^{p+1-n}\nabla_T\mathcal{U}_n) \simeq
R^{p-n}\Omega^u_{n-1}\,.
\end{align}
where on the RHS~$\Omega^{u}_{n-1}$ depends on the
functions~$\{G_{m,k},B_{m,k},U_{m,k},m_{u,1}\}$, for~$m\in[1,n-1]$
and~$k\in[0,N_n^\phi]$, and their derivatives.
Also,~$\Omega^{u}_{0}:=0$. We can split~$\Omega^{u}_{n-1}$ in the
following way,
\begin{align}\label{expH}
\Omega^{u}_{n-1} = \sum_{i=0}^{N_n^{\Omega^u}}(\log
R)^i\Omega^{u}_{n-1,i}(\psi^*)\,.
\end{align}
It is worth noting that the specific form of~$\Omega^{u}_{n-1,i}$ has
no influence on the proof of our hypothesis, as long as it is possible
to write~\eqref{expH}. We can now integrate~\eqref{umot3} in order to
get the asymptotic behavior of~$\mathcal{U}_n$ in terms
of~$\{G_{m,k},B_{m,k},U_{m,k},m_{u,1}\}$. Plugging~\eqref{expH}
into~\eqref{umot3},
\begin{align}
  \nabla_\psi(R^{p+1-n}\nabla_T\mathcal{U}_n) \simeq R^{p-n}
  \sum_{i=0}^{N_n^{\Omega^u}}(\log
  R)^i\Omega^{u}_{n-1,i}(\psi^*)\,,\non
\end{align}
and integrating it along integral curves of~$\psi^a$ and~$\p_T^a$
gives, once more, two different results. For~$n\neq p+1$,
\begin{align}\label{ugeneral1}
  \mathcal{U}_n \simeq& \sum_{i=0}^{N_n^{\Omega^u}}(\log R)^i
  \sum_{j=i}^{N_n^{\Omega^u}}
  -\frac{1}{(n-1-p)^{j-i+1}}\frac{j!}{i!}\int_{T_0}^T
  \Omega^{U}_{n-1,j}dT'\non\\ &+ m_{u,n}\non\\ =&
  \sum_{i=0}^{N_n^{\Omega^u}} (\log R)^i U_{n,i}(\psi^*)\,,
\end{align}
whereas for~$n=p+1$,
\begin{align}\label{ugeneral2}
  \mathcal{U}_n \simeq& \sum_{i=0}^{N_n^{\Omega^u}}\frac{(\log
    R)^{i+1}}{i+1} \int_{T_0}^T \Omega^{U}_{n-1,i}dT'+
  m_{u,n}\non\\ =& \sum_{i=0}^{N_n^{\Omega^u}+1} (\log R)^i
  U_{n,i}(\psi^*)\,.
\end{align}
By induction,
\begin{align}
  u = \frac{m_{u,1}}{R}+\sum_{n=2}^{\infty}\sum_{k=0}^{N_n^u}
  \frac{(\log R)^k U_{n,k}(\psi^*)}{R^n}\,.
\end{align}
This concludes the proof. These results can be packaged in the
following theorem.
\begin{thm} \label{T2} Let there be any number of evolved fields, each
  satisfying one of the following 3 wave equations,
\begin{align}\label{gbugeneral2}
  &\mathring{\square} g = \mathcal{N}_g\,,\non\\ &\mathring{\square}
  b = -\frac{1}{2}(\nabla_T g_1)^2-\frac{1}{2}(\nabla_T g_2)^2 +
  \mathcal{N}_b\,,\non\\ &\mathring{\square} u = \tfrac{2p}{R}\nabla_T
  u + \mathcal{N}_u\,,
\end{align}
where~$g_{ab}$ is an asymptotically flat metric, $g_1$ and~$g_2$
satisfy the first equation in~\eqref{gbugeneral}, and~$p$ may be
different for different functions. Fields that satisfy the equations
for~$g$, $b$ and~$u$ admit polyhomogeneous expansions near null
infinity of the types,
\begin{align}\label{polyhomogeneral}
  &g = \frac{G_{1,0}(\psi^*)}{R} +
  \sum_{n=2}^{\infty}\sum_{k=0}^{N_n^g} \frac{(\log R)^k
    G_{n,k}(\psi^*)}{R^n}\non\\ &b = \frac{B_{1}}{R} +
  \sum_{n=2}^{\infty}\sum_{k=0}^{N_n^b} \frac{(\log R)^k
    B_{n,k}(\psi^*)}{R^n}\non\\ &u =
  \frac{m_{u,1}}{R}+\sum_{n=2}^{\infty}\sum_{k=0}^{N_n^u} \frac{(\log
    R)^k U_{n,k}(\psi^*)}{R^n} \,,
\end{align}
respectively, with initial data on~$\mathcal{S}$ of the type,
\begin{align}
  \begin{cases}
    g\rvert_{\mathcal{S}}=
    \sum_{n=1}^{\infty}\frac{m_{g,n}}{R^n}\\ b\rvert_{\mathcal{S}}=
    \sum_{n=1}^\infty \frac{m_{b,n}}{R^n}\\ u\rvert_{\mathcal{S}}=
    \sum_{n=1}^\infty \frac{m_{u,n}}{R^n}
  \end{cases}\,,
  \begin{cases}
    \nabla_Tg\rvert_{\mathcal{S}}=
    O_{\mathcal{S}}(R^{-2})\\ \nabla_Tb\rvert_{\mathcal{S}}=
    O_{\mathcal{S}}(R^{-2})\\ \nabla_Tu\rvert_{\mathcal{S}}=
    O_{\mathcal{S}}(R^{-2})
  \end{cases}\,,
\end{align}
where~$m_{\phi,n}$ are scalar functions that are independent of~$T$
and~$R$. This is valid outside a compact ball centered at~$R=0$.
\end{thm}
\begin{remark}\label{remark1} Consider a system in which all of the
  evolved fields satisfy either the good equation or the ugly one.  We
  know that the good equation cannot create logs, it can only inherit
  them though coupling with the other equations.  Since there is no
  bad field in such a system, the only logs that may appear are the
  ones created by the ugly equation.  Let us also assume that all the
  ugly fields have the same~$p$. In that case, for all~$n<p+1$, there
  can be no logs in~\eqref{expH}, or in other words,
  $N_n^{\Omega^u}=0$. This implies that up to order~$p+1$,
\begin{align}
	\mathcal{U}_n \simeq U_{n,0}(\psi^*)\,.
\end{align}
However, when we get to order~$n=p+1$, because the integration
generates one more power of~$\log R$ than what already existed, in our
system this is the first order at which a log may appear. For~$n>p+1$,
integration along~$\psi$ no longer generates higher powers of logs, so
any increase in~$N_n^{u}$ may only come from nonlinearities. So our
system admits the following polyhomogeneous expansion,
\begin{align}\label{polyhomosuperuglies}
  &g = \frac{G_{1,0}(\psi^*)}{R} +\sum_{n=2}^{p}
  \frac{G_{n,0}(\psi^*)}{R^n}+ \sum_{n=p+1}^{\infty}\sum_{k=0}^{N_n^g}
  \frac{(\log R)^k G_{n,k}(\psi^*)}{R^n}\non\\ &u =
  \frac{m_{u,1}}{R}+\sum_{n=2}^{p}
  \frac{U_{n,0}(\psi^*)}{R^n}+\sum_{n=p+1}^{\infty}\sum_{k=0}^{N_n^u}
  \frac{(\log R)^k U_{n,k}(\psi^*)}{R^n}\,.
\end{align}
In other words, in a system with the properties described above, the
polyhomogeneous expansions can have no logs up to order~$p+1$. This
result will be critical to our analysis of peeling.

\end{remark}

\section{Reduced Einstein field equations}\label{EFE}

In this section we derive the field equations of GR in GHG using our
preferred variables. In this work we are primarily concerned with the
asymptotic properties of solutions, which we obtain by brute-force
integration, with many terms turning out to be irrelevant. Therefore
we do not focus on obtaining the cleanest possible geometric
derivation.

A concise way to find some of the wave equations is by computing the
wave equation for the null covector~$\sigma_a$ and commuting
derivatives to get some component of the Ricci
tensor. From~\eqref{eq:xitoeta} we get,
\begin{align*}
\square \sigma_a = & -\square \mathcal{C}_+^R\nabla_a T+\square
\mathcal{C}_A^+\nabla_a
\theta^A-2g^{bc}\nabla_b\mathcal{C}_+^R\nabla_c\nabla_aT \non
\\ &+2g^{bc}\nabla_b\mathcal{C}_A^+\nabla_c\nabla_a\theta^A -
\mathcal{C}_+^Rg^{bc}\nabla_b\nabla_c\nabla_a T \non\\ &+
g^{bc}\nabla_b\nabla_c\nabla_a R+
\mathcal{C}_A^+g^{bc}\nabla_b\nabla_c\nabla_a \theta^A\,.
\end{align*}
Commuting the Levi-Civita covariant derivatives~$\nabla$ and using the
Ricci identity, reveals
\begin{align}
  \square \sigma_a = & R_{a\sigma} -\square
  \mathcal{C}_+^R\nabla_a T+\square \mathcal{C}_A^+\nabla_a
  \theta^A \non\\
  &-2g^{bc}\nabla_b\mathcal{C}_+^R\nabla_c\nabla_aT
  +2g^{bc}\nabla_b\mathcal{C}_A^+\nabla_c\nabla_a\theta^A
  \non\\
  &- \mathcal{C}_+^R\nabla_a\square T + \nabla_a\square
  R+ \mathcal{C}_A^+\nabla_a\square \theta^A\,.
\end{align}
Replacing the first derivatives of metric functions
with~\eqref{ConnectionComps}, using the fact that,
\begin{align}
 &\square T = -\Gb^a\nabla_aT\,,\\
 &\square R = -\Gb^a\nabla_aR\,,\\
 &\square \theta^A = -\Gb^a\nabla_a\theta^A\,,
\end{align}
and writing~$\nabla_aT$ in terms of the null covectors
with~\eqref{eq:xiinshellchart},
\begin{align}
  \tau
  \nabla_aT=-\sigma_a-\sigmab_a+\mathcal{C}_A^+
  \nabla_a\theta^A+\mathcal{C}_A^-\nabla_a\theta^A\,,
\end{align}
we get the expression,
\begin{align} \label{boxsigma}
  \square \sigma_a = & R_{a\sigma} -\square
  \mathcal{C}_+^R\nabla_a T+\square \mathcal{C}_A^+\nabla_a
  \theta^A - \sigma_b\nabla_a\Gb^b + \Gb^b\Gb_a{}^\sigma{}_b\non\\
  &+\frac{2}{\tau}\Gb_\psi{}^\sigma{}_b(\Gb_a{}^{\sigma
    b}+\Gb_a{}^{\sigmab b})-2\Gb_{bAa}\Gb^{(\sigma A)b}\,.
\end{align}
We can contract this equation with~$\psi^a$, $\psib^a$ and~$\sg_b{}^a$
to obtain wave equations for~$\mathcal{C}_+^R$, $\varphi$
and~$\mathcal{C}_A^+$, respectively, and we can contract the analogous
equation for~$\square \sigmab_a$ with~$\psib^a$ and~$\sg_b{}^a$ to
obtain wave equations for~$\mathcal{C}_-^R$
and~$\mathcal{C}_A^-$. First we introduce the reduced Ricci tensor
\begin{flalign}\label{eq:ReducedRicciDefinition}
  \mathcal{R}_{ab} := R_{ab} - \nabla_{(a}Z_{b)}+ W_{ab} \; \text{with} \; Z^a :=
 \mathbullet{\Gamma}^a + F^a &&
\end{flalign}
where~$\Gb^a:=g^{bc}\Gb_b{}^a{}_c$, $F^a$ are the gauge source
functions and~$W_{ab}$ denotes a generic constraint addition where by
generic we mean any expression for which the propagation of the
constraints construction holds.  It can be shown, see for
instance~\cite{GasHil18}, that as long as~$W_{ab}$ is a homogeneous
expression in~$Z^a$, i.e. $Z^a=0 \implies W_{ab}=0$, the constraints
propagate. In other words, if~$Z^a=0$ and~$\nabla_aZ^b=0$ on a
spacelike hypersurface~$\Sigma \subset \mathcal{M}$ then~$Z^a=0$ in
the future domain of dependence~$\mathcal{D}^+(\Sigma)$. For
conciseness we also introduce the tensor~$\tilde{\mathcal{R}}_{ab}$
defined as,
\begin{flalign}\label{Rtilde}
  \tilde{\mathcal{R}}_{ab} :=
  \mathcal{R}_{ab} + \nabla_{(a}F_{b)} - W_{ab}\,.
\end{flalign}
Now we contract~\eqref{boxsigma} with $\psi^a$ to obtain,
\begin{align}\label{psiboxsigma}
  \psi^a\square \sigma_a = &
  e^{-\varphi}\tilde{\mathcal{R}}_{\psi\psi}
  -\mathbullet{\square}
  \mathcal{C}_+^R+\frac{2}{\tau}\Gb_\psi{}^\sigma{}_a(\Gb_\psi{}^{\sigma
    a}+\Gb_\psi{}^{\sigmab
    a})\non\\
  &-4\sg_{bc}\Gb_\psi{}^b{}_a\Gb^{(\sigma c)a}\,.
\end{align}
Note that all derivatives of~$\Gb_a{}^b{}_c$ canceled and also that
we have replaced~$\square\mathcal{C}_+^R$ with~$\mathbullet{\square}
\mathcal{C}_+^R$. On the other hand, a direct computation yields,
\begin{align}\label{psiboxsigmaLHS}
  \psi^a\square \sigma_a =
  -e^{-\varphi}\sg_{bc}\Gb_\psi{}^b{}_a\Gb_\psi{}^c{}^a\,.
\end{align}
We can now equate the LHS of equations~\eqref{psiboxsigma}
and~\eqref{psiboxsigmaLHS}, and solve for~$\mathbullet{\square}
\mathcal{C}_+^R$. This gives,
\begin{align}\label{eqCpR}
  \mathbullet{\square} \mathcal{C}_+^R =&
  e^{-\varphi}\tilde{\mathcal{R}}_{\psi\psi} +
  \frac{2}{\tau}\Gb_\psi{}^\sigma{}_a\Gb_\psi{}^{\sigma
    a}-e^{-\varphi}\Gb_\psi{}^b{}_a\Gb_{\psi{}b}{}^a \\
  & -2\sg_a{}^c\Gb_\psi{}^a{}_b\Gb_c{}^{\sigma b}\non\,.
\end{align}
If we contract~\eqref{boxsigma} instead with~$\psib^a$ we get,
\begin{align}\label{psibarboxsigma}
  &\psib^a\square \sigma_a =
  e^{-\varphi}\tilde{\mathcal{R}}_{\psib\psi} -
  e^{-\varphi}\psi^a\nabla_{\psib}\Gb_a +
  e^{-\varphi}\psib^b\psi^d\nabla_{(d}\Gb_{b)}- \square
  \mathcal{C}_+^R\non\\
  &+\frac{2}{\tau}\Gb_{\psi}{}^\sigma{}_a\left(\Gb_{\psib}{}^{\sigma
    a}+\Gb_{\psib}{}^{\sigmab a}\right) -
  2\sg_{de}\Gb_{\psib}{}^d{}_c\left(\Gb^{\sigma ec}+\Gb^{e\sigmab
    c}\right)
\end{align}
Computing the LHS without commutation gives,
\begin{align}\label{psibarboxsigmaLHS}
	\psib^a\square \sigma_a = &
        -e^{-\varphi}\tilde{\mathcal{R}}_{\psib\psi} +
        e^{-\varphi}\psib^a\nabla_\psi\Gb_a -
        e^{-\varphi}\psib^b\psi^d\nabla_{(d}\Gb_{b)}\non\\ &
        \frac{1}{\tau}\left(\Gb_{\psib}{}^\sigma{}_a\Gb_{\psib}{}^{\sigma
          a}+2\Gb_{\psib}{}^\sigma{}_a\Gb_{\psi}{}^{\sigma
          a}-\Gb_{\psi}{}^{\sigmab}{}_a\Gb_{\psi}{}^{\sigmab
          a}\right)\non\\ & - \square \mathcal{C}_+^R+\tau\square
        \varphi - \Gb^a\Gb_\psi^{\sigmab}{}_a\,,
\end{align}
where, in order to eliminate terms
like~$g^{cd}\nabla_c\mathbullet{\Gamma}^a{}_{bd}$, we have used the
following identity,
\begin{align}
  \label{eq:DerGammaRicciIdentity}
  g^{cd}\nabla_c\mathbullet{\Gamma}^a{}_{bd} =
  \mathbullet{\Gamma}^{acd}\mathbullet{\Gamma}_{cbd}-R_{b}{}^a +
  \nabla_b\Gamma^a\,.
\end{align}
Putting the two sides together cancels all derivatives
of~$\Gb_a{}^b{}_c$. Solving for~$\mathbullet{\square}\varphi$ finally
gives,
\begin{align}\label{eqvarphi}
  \mathbullet{\square} \varphi =&
  \frac{2}{\tau}\tilde{\mathcal{R}}_{\psi}{}^{\sigmab} -
  \frac{2}{\tau^2}\left[\Gb_{\psib}{}^a{}_{(\psi}\Gb_{\psib)}{}^{\sigma}{}_a
    +
    \Gb_{\psi}{}^a{}_{(\psib}\Gb_{\psi)}{}^{\sigmab}{}_a\right]\\ &
  +\Gb_a{}^{db}\Gb^{ac}{}_e\left[2\sg_{cb}(\delta_d^e-\sg_d{}^e)
    +\sg_{cd}(\delta_b^e-\sg_b{}^e)\right]\,.\non
\end{align}
The contraction of~\eqref{boxsigma} with~$\sg_b{}^a$ yields,
\begin{align}\label{gslashboxsigma}
  &\sg_A{}^a\square \sigma_a =
  e^{-\varphi}\sg_A{}^a\tilde{\mathcal{R}}_{a\psi} +
  \frac{\mathcal{C}^+_A+\mathcal{C}^-_A}{\tau}\mathbullet{\square}
  \mathcal{C}_+^R + \mathbullet{\square} \mathcal{C}_A^+\non \\ &+
  \frac{2}{\tau}\sg_A{}^a\Gb_\psi{}^\sigma{}_b\left(\Gb_a{}^{\sigma
    b}+\Gb_a{}^{\sigmab b}\right) -
  4\sg_A{}^a\sg_{de}\Gb_a{}^d{}_b\Gb^{(\sigma
    e)b}\non\\
  &-e^{-\varphi}\sg_A{}^a\psi^b\nabla_{[a}\Gb_{c]}
  -e^{-\varphi}\sg_{aA}\Gb^c\Gb_\psi{}^a{}_c\,.
\end{align}
On the other hand, the LHS gives,
\begin{align}\label{gslashboxsigmaLHS}
  &\sg_A{}^a\square \sigma_a =
  -e^{-\varphi}\sg_A{}^a\tilde{\mathcal{R}}_{a\psi}
  -e^{-\varphi}\sg_A{}^a\psi^b\nabla_{[a}\Gb_{c]}\non \\
  &-e^{-\varphi}\sg_{aA}\Gb^c\Gb_\psi{}^a{}_c
  +\frac{2}{\tau}\sg_{Ab}\Gb_\psi{}^b{}_a\left(\Gb_{\psib}{}^{\sigma
    a}-\Gb_{\psi}{}^{\sigma a}\right) \non \\
  &+ 2e^{-\varphi}\sg_{Ab}\sg_c{}^e\Gb_a{}^b{}_e\Gb_\psi{}^{ca}\,.
\end{align}
Putting~\eqref{gslashboxsigma} and~\eqref{gslashboxsigmaLHS} together
yields,
\begin{align}\label{eqCpA}
	&\mathbullet{\square} \mathcal{C}^+_A =
  -e^{-\varphi}\sg_A{}^a\tilde{\mathcal{R}}_{a\psi} +
  \frac{\mathcal{C}^+_A+\mathcal{C}^-_A}{\tau}\mathbullet{\square}
  \mathcal{C}_+^R\non \\ &+ \frac{2}{\tau}e^{-\varphi}\Gb_{\psi
    A}{}^{a}(\Gb_{\psib}{}^\sigma{}_a-\Gb_{\psi}{}^\sigma{}_a) -
  \frac{2}{\tau}\Gb_\psi{}^{\sigma}{}_b(\Gb_A{}^{\sigma
    b}+\Gb_A{}^{\sigmab b})\non\\ & +2e^{-\varphi}\sg_f{}^e\Gb_{a A
    e}\Gb_\psi{}^{fa} -4\sg_{de}\Gb_A{}^d{}_b\Gb^{(\sigma e) b}\,.
\end{align}
A completely analogous calculation using the
contractions~$\psib^a\square \sigmab_a$ and~$\sg_A{}^a\square
\sigmab_a$ allows us to find wave equations for~$\mathcal{C}_-^R$
and~$\mathcal{C}_A^-$, respectively.
\begin{align}\label{eqCmR}
  \mathbullet{\square} \mathcal{C}_-^R =&-
  e^{-\varphi}\tilde{\mathcal{R}}_{\psib\psib} -
  \frac{2}{\tau}\Gb_{\psib}{}^{\sigmab}{}_a\Gb_{\psib}{}^{{\sigmab}
    a}+e^{-\varphi}\Gb_{\psib}{}^b{}_a\Gb_{\psib{}b}{}^a \\
  &+2\sg_a{}^c\Gb_{\psib}{}^a{}_b\Gb_c{}^{\sigmab b}\non\,,
\end{align}
and,
\begin{align}\label{eqCmA}
  &\mathbullet{\square} \mathcal{C}^-_A =
  -e^{-\varphi}\sg_A{}^a\tilde{\mathcal{R}}_{a\psib} -
  \frac{\mathcal{C}^+_A+\mathcal{C}^-_A}{\tau}\mathbullet{\square}
  \mathcal{C}_-^R\non \\
  &+ \frac{2}{\tau}e^{-\varphi}\Gb_{\psib
    A}{}^{a}(\Gb_{\psi}{}^{\sigmab}{}_a-\Gb_{\psib}{}^{\sigmab}{}_a) -
  \frac{2}{\tau}\Gb_{\psib}{}^{\sigmab}{}_b(\Gb_A{}^{\sigmab
    b}+\Gb_A{}^{\sigma b})\non\\
  & +2e^{-\varphi}\sg_f{}^e\Gb_{a A e}
  \Gb_{\psib}{}^{fa} -4\sg_{de}\Gb_A{}^d{}_b\Gb^{(\sigmab e) b}\,.
\end{align}
To obtain the final 3 equations associated to the angular sector of
the metric~$g_{ab}$: $\epsilon$, $h_+$ and~$h_\times$, we recall the
second Cartan structure equation, which in our notation reads,
\begin{align}\label{eq:SecondCartanStructure}
  R_{abc}{}^{d}=2g^{ef}\mathbullet{\Gamma}^{d}{}_{[a|e}\mathbullet{\Gamma}_{f|b]c}
  -2\nabla_{[a}\mathbullet{\Gamma}^{d}{}_{b]c}.
\end{align}
Tracing and projecting equation~\eqref{eq:SecondCartanStructure}
appropriately gives,
\begin{align}
  \slashed{g}^{ab}R_{ab} =
  \mathbullet{\Gamma}^{abd}\mathbullet{\Gamma}_{bd}{}^{c}\slashed{g}_{ac}
  -\mathbullet{\Gamma}^{a}\mathbullet{\Gamma}^{b}{}_{a}{}^{c}\slashed{g}_{bc}
  + \slashed{g}^{ab}\nabla_b \mathbullet{\Gamma}_{a}
  -\slashed{g}^{ab}\nabla_c \mathbullet{\Gamma}_{ab}{}^{c}.
\end{align}
A calculation using~\eqref{ConnectionComps} renders the following wave
equation for~$\epsilon$,
\begin{align}\label{eqeps}
  \mathbullet{\square} (\epsilon + \mathbullet{\epsilon}) &=
  -\tilde{\slashed{\mathcal{R}}}
  -\Gb^{bac}\Gb_e{}^d{}_c
  \left[\sg_{ad}(\delta_b^e-\sg_b{}^e)-\sg_{bd}\sg_a{}^e\right]\,,
\end{align}
To get the remaining two wave equations we
trace~\eqref{eq:SecondCartanStructure} again, raise its indices and
use~\eqref{Rtilde} to obtain
\begin{align}\label{CartanUpstairs}
  \tilde{\mathcal{R}}^{ab}=
  -\nabla_c\Gb^{(ab)c}-\Gb^c\Gb^{(ab)}{}_c+\Gb_d{}^{c(a}\Gb_c{}^{b)d}\,.
\end{align}
We need to eliminate the first term on the RHS and to do that we take a
Shell derivative of the inverse metric,
\begin{align}
  \mbn_cg^{ab} = -g^{ad}g^{be}\mbn_cg_{de} = -2\Gb^{(ab)}{}_c\,,
\end{align}
then we take a divergence on the downstairs index with the Levi-Civita
covariant derivative and change that derivative to a $\mbn$ with the
appropriate transition tensor,
\begin{align}\label{nablaChristToboxg}
  -2\nabla_c\Gb^{(ab)c} =& \mathbullet{\square}g^{ab} +
  2\Gb^c\Gb^{(ab)}{}_c - 2\Gb^{ca}{}_d\Gb~^{(db)}{}_c \non\\
  &-2\Gb^{cb}{}_d\Gb~^{(da)}{}_c\,.
\end{align}
We can now replace~\eqref{nablaChristToboxg} in~\eqref{CartanUpstairs}
to get,
\begin{align}\label{CartanUpstairsFinal}
	\tilde{\mathcal{R}}^{ab}=
        \frac{1}{2}\mathbullet{\square}g^{ab}+\Gb_c{}^{a}{}_d\Gb^{cbd}\,.
\end{align}
On the other hand, a rather long calculation
using~\eqref{eq:qconformaltransformation} gives,
\begin{align}\label{boxgslash}
	&\mathbullet{\square}\sg^{AB} =
  \frac{e^{-\epsilon}}{R^2}\mathbullet{\square}(q^{-1})^{AB}\non
  \\ &-\sg^{AB}\left[\mathbullet{\square} (\epsilon +
    \mathbullet{\epsilon}) - \sg^{\theta\theta}\frac{1+\cos^2
      \theta}{\sin^2 \theta}
    +2\cot\theta\sg_e{}^d\sg^{c\theta}\Gb_d{}^e{}_c
    \right]\non\\ &-4\sg_f{}^A\sg_e{}^B\Gb^{(ef)}{}_c
  \left[\cot\theta\sg^{c\theta}- \sg_e{}^d\Gb_d{}^{ec}\right]\,.
\end{align}
Taking only the angular components of
equation~\eqref{CartanUpstairsFinal}, we can substitute the first term
on the RHS with~\eqref{boxgslash} and project the whole equation with,
\begin{align*}
  \slashed{S}_{a}{}^{c}{}_{b}{}^{d}=\slashed{g}_{a}{}^{(c}\slashed{g}_{b}{}^{d)}
  -\tfrac{1}{2}\slashed{g}_{ab}\slashed{g}^{cd}\,.
\end{align*}
Finally, for the two remaining metric functions~$h_+$ and~$h_\times$,
we get the concise expression,
\begin{align}\label{eqq}
  \slashed{S}_A{}^c{}_B{}^d\mathbullet{\square} q^{AB}
  =&2R^2e^{\epsilon}\slashed{S}_A{}^c{}_B{}^d\left[\tilde{\mathcal{R}}^{AB}+
    \Gb_e{}^A{}_f\Gb^{eBf}\right. \\ &\left.-
    2\Gb^{(AB)}{}_c\left(\sg_f{}^e\Gb_e{}^{fc}-\cot \theta
    \sg^{c\theta}\right)\right]\non\,.
\end{align}
With this we conclude the derivation of the reduced EFE, a system of
non-linear wave equations, one for each of the 10 metric
components,~\eqref{eqCpR}, \eqref{eqvarphi}, \eqref{eqCpA},
\eqref{eqCmR}, \eqref{eqCmA}, \eqref{eqeps} and~\eqref{eqq}. The next
step in this analysis is to see how to identify each field as {\it
  good}, {\it bad} or {\it ugly} and we explore that in the next
section.

\section{Asymptotics in Cartesian Harmonic Gauge}\label{AsCartHarm}

The aim of this section is to find the functional form of the first
few orders of the polyhomogeneous expansions for our metric variables
near null infinity in GR. In other words, we want to use the tools of
the last sections to discover the maximum power of~$\log R$ at each
order~$n$ for each metric function~$\phi$. To do that we begin by
finding the terms that can contribute to leading order, then we verify
that each field can be classified as a {\it good}, a {\it bad} or an
{\it ugly}. Finally, we use Theorem~\ref{T2} to show that these fields
admit a polyhomogeneous expansion and we integrate the wave equations
to find the maximum power of~$\log R$, $N_n^\phi$, that is allowed. To
illustrate the fact that the information about the functional form of
a field~$\phi$ to some order~$n$ is contained in the set of
integers~$N_n^\phi$, we consider an example: Let~$\phi$ have the
functional form,
\begin{align}
	&N_1^\phi = 0\non\quad,\quad N_2^\phi = 2\non\,,
\end{align}
then we can write its first~$2$ orders as,
\begin{align}
	\phi \simeq \frac{\Phi_{1,0}}{R} + \frac{\Phi_{2,0}+\log
          R\Phi_{2,1}+(\log R)^2\Phi_{2,2}}{R^2}\,,\non
\end{align}
where the functions $\Phi_{n,i}=\Phi_{n,i}(\psi^*)$.

\paragraph*{Cartesian harmonic gauge:}
Observe that up to this point the gauge source functions
in~$\tilde{\mathcal{R}}_{ab}$ have not been specified. Let,
\begin{align}\label{Fgeneral}
F^a = \ring{F}^a\,,
\end{align}
where,
\begin{align}\label{eq:GaugeSourcesForCartesianHarmonic}
  \ring{F}^a =
  g^{bc}\Gamma[\mathring{\nabla},\mathbullet{\nabla}]^{a}{}_{bc}\,.
\end{align}
Notice that this choice implies,
using~\eqref{eq:CartesianAndShellConnectionRelation}
and~\eqref{eq:ReducedRicciDefinition},
that~$g^{bc}\Gamma[\nabla,\mathring{\nabla}]_b{}^a{}_c=0$. In other
words, $F^{a}=\ring{F}^a$ implies that the Cartesian
coordinates~$X^{\ul{\mu}}$ are harmonic --- recall
that~$\Gamma[\mathring{\nabla},\mathbullet{\nabla}]_{a}{}^{b}{}_{c}$
are given functions of the coordinates. For clarity, we write down the
four components of~$\ring{F}^a$ explicitly,
\begin{align}
  &\ring{F}^\sigma =
  \frac{2e^{-\epsilon}}{R}\cosh h_+\cosh h_\times
  + \mathcal{C}^+_A\ring{F}^A\,,\non\\
  &\ring{F}^{\sigmab} =
  -\frac{2e^{-\epsilon}}{R}\cosh h_+\cosh h_\times
  + \mathcal{C}^-_A\ring{F}^A\,,\non\\
  &\ring{F}^\theta =
  \frac{e^{-\epsilon}\cot \theta}{R^2}e^{h_+}\cosh h_\times
  -\frac{2}{R}\sg^{R\theta}\,,\non\\
  &\ring{F}^\phi =
  -\frac{2e^{-\epsilon}\cot \theta}{\sin \theta R^2}\sinh h_\times
  -\frac{2}{R}\sg^{R\phi}\,.
\end{align}

\paragraph*{Constraint addition:} We want to be able to
separate all of the 10 metric functions into \textit{goods},
\textit{bads} and \textit{uglies} in the same way as
in~\cite{DuaFenGasHil21}. As it turns out it is possible to write four
of the metric functions as uglies,
\begin{align}
  \mathring{\square}\phi = \frac{2}{R}\nabla_T\phi + \mathcal{N}_\phi\,,
\end{align}
through constraint addition, where~$\mathcal{N}_\phi$ is a stratified
null form and~$\phi$ is a metric function. To see this, we need to
compute the constraints~$Z^a$ to leading order and add specific
constraint terms encoded in~$W_{ab}$ to some of the wave equations.
Asymptotically, the constraints are the following,
\begin{align}
  &Z^\sigma = 2\nabla_T \mathcal{C}_+^R+o^+(1)\,,\non\\
  &Z^{\sigmab} = -2\nabla_T \epsilon+o^+(1)\,,\non\\
  &Z^\theta = -\frac{1}{R^2}\nabla_T \mathcal{C}_\theta^++o^+(R^{-2})\,,\non\\
  &Z^\phi = -\frac{1}{R^2\sin^2\theta}\nabla_T
  \mathcal{C}_\phi^++o^+(R^{-2})\,.
\end{align}
As each of the components of~$Z^a$ involves a~$\nabla_T$ derivative of
a specific metric function, we can choose the components of~$W_{ab}$
in order to turn~$\mathcal{C}_+^R$, $\epsilon$ and~$\mathcal{C}_A^+$
into ugly fields. One choice of components that does this, and this
choice is highly non-unique, is the following,
\begin{align}\label{ConstAdd}
  &W_{\psi\psi} =
  -\frac{1}{2}(Z^\sigma)^2+\frac{1}{R}Z^\sigma\,,\non\\ &\slashed{W} =
  -\frac{1}{R}Z^{\sigmab}\,,\non\\ &W_{\psi A} = \frac{2}{R}Z^A\,,
\end{align}
with all remaining components set to zero. Note that the indices in
the last equation are not incorrect, since~$A$ here is not a tensorial
index, but a label.

\paragraph*{First order asymptotic system:} In order
to classify all wave equations as {\it good}, {\it bad} or {\it ugly},
we need to rescale~$\mathcal{C}_A^\pm$ to account for the~$R$ factor
in~\eqref{asympflatness}. Therefore we define,
\begin{align}
	R\hat{\mathcal{C}}_A^\pm = \mathcal{C}_A^\pm\,.
\end{align}
Using Cartesian harmonic gauge, the constraint
additions~\eqref{ConstAdd} and the wave equations~\eqref{eqCpR},
\eqref{eqvarphi}, \eqref{eqCpA}, \eqref{eqCmR},
\eqref{eqCmA},~\eqref{eqeps} and~\eqref{eqq}, the system takes the
following form,
\begin{align}
 &  \mathring{\square} \varphi = \mathcal{N}_\varphi\,,\non \\
   & \mathring{\square} \mathcal{C}_{+}^R  =
   \frac{2}{R}\nabla_T\mathcal{C}_+^R+
   \mathcal{N}_{\mathcal{C}_{+}^R} \,,\non\\
   &\mathring{\square} \mathcal{C}_{-}^R  =
   - \frac{1}{2}(\nabla_T h_+)^2-\frac{1}{2}(\nabla_T h_\times)^2
   + \mathcal{N}_{\mathcal{C}_{-}^R}\,, \non\\
   &\mathring{\square} \hat{\mathcal{C}}_A^+
   = \frac{2}{R}\nabla_T\hat{\mathcal{C}}_A^+
   +\mathcal{N}_{\mathcal{C}_A^+} \,,\non\\
   &\mathring{\square} \hat{\mathcal{C}}_A^- =
   \frac{4}{R}\nabla_T\hat{\mathcal{C}}_A^-
   +\mathcal{N}_{\mathcal{C}_A^-}\,,\non \\
   &\mathring{\square} \epsilon  =
   \frac{2}{R}\nabla_T\epsilon +\mathcal{N}_{\epsilon}\,,\non\\
   &\mathring{\square} h_+ = \mathcal{N}_{h_+}\,,\non \\
   &\mathring{\square} h_\times = \mathcal{N}_{h_\times}\,.
 \end{align}
We can easily see that the fields~$\varphi$, $h_+$ and~$h_\times$
satisfy the good equation, whereas~$\mathcal{C}_-^R$ satisfies the bad
one, $\mathcal{C}_+^R$, $\hat{\mathcal{C}}_A^+$ and~$\epsilon$ satisfy
the ugly equation with~$p=1$ and~$\hat{\mathcal{C}}_A^-$ the ugly
equation with~$p=2$. This already gives us the functional form of the
fields to first order in~$R^{-1}$, 
\begin{align}\label{FunctionalFormCart1}
  &N_1^{\varphi}=N_1^{\mathcal{C}_+^R}=N_1^{\mathcal{C}_A^\pm}=N_1^{\epsilon}
  =N_1^{h_+}=N_1^{h_\times}=0\,,\non\\
  &N_1^{\mathcal{C}_-^R}=1\,.
\end{align}

\paragraph*{Second and third order:} We can now apply Theorem~\ref{T2}
to assert that GR in Cartesian harmonic gauge and with constraint
additions~\eqref{ConstAdd} admits a polyhomogeneous expansion of the
type~\eqref{gbugeneral}.  In order to find the functional form for
second order, we plug~\eqref{gbugeneral} into the wave equations and
formally equate terms proportional to~$R^{-3}$, we put all terms
containing~$\Phi_2$ on the LHS and all the rest we collect on the RHS
and name
it~$\Omega_1^{\phi}$. Here~$\Phi_n:=\sum_{k=0}^{N_n^\phi}(\log
R)^k\Phi_{n,k}(\psi^*)$, where~$\Phi_{n,k} =
\{G_{n,k},B_{n,k},U_{n,k}\}$. Note that we are not interested in the
exact dependence of~$\Omega_1^{\phi}$ on the fields, but we {\it are}
interested in its functional form. Integrating the resulting equations
we find that the functional form to second order is given by,
\begin{align}\label{FunctionalFormCart2}
  &N_2^{\varphi}=N_2^{\mathcal{C}_+^R}
  =N_2^{\mathcal{C}_A^+}=N_2^{\epsilon}=N_1^{h_+}=N_2^{h_\times}=1\,,\non\\
	&N_2^{\mathcal{C}_-^R}=N_2^{\mathcal{C}_A^-}=2\,.
\end{align}
Equating terms of order~$R^{-4}$ and following exactly the same
procedure we find the functional form to third order,
\begin{align}\label{FunctionalFormCart3}
  &N_3^{\mathcal{C}_+^R}=1\,,\non\\
  &N_3^{\varphi}=N_3^{\epsilon}=N_3^{\mathcal{C}_A^+}
  =N_3^{h_+}=N_3^{h_\times}=2\,,\non\\
  &N_3^{\mathcal{C}_-^R}=N_3^{\mathcal{C}_A^-}=3\,.
\end{align}
Given the increasing complexity and computational time these
expansions demand, we stop at third order, since that is already
enough to show the result of the next section.

The full calculations needed to arrive at these results are given in
the Mathematica notebooks that accompany the paper. They rely heavily
on the xAct~\cite{xAct_web} package.

\section{Violation of peeling}\label{NoPeeling}

As stated above, smooth null infinity implies the satisfaction of
peeling, i.e. that the components of the Weyl curvature tensor fall
off with certain negative integer powers of radius. We will see here
that our choice of gauge~\eqref{Fgeneral}, together with our
constraint addition~\eqref{ConstAdd} gives rise to a violation of
peeling by introducing powers of~$\log R$ in the leading order terms
of some of the components. The Weyl tensor has 10 independent
components which can be described concisely by the \textit{Weyl
  scalars}, 5 complex numbers that result from the contraction of the
Weyl tensor with unit null vectors.  Following the conventions
of~\cite{Alc08}, we define the null tetrad (with normalization~$ l_a
n^a = - m_a \bar{m}^a = -1 $):
\begin{align}
  &l^a=\frac{\psi^a}{\sqrt{\tau \, e^\varphi}} \,,\non\\
  &n^a=\frac{\ul{\psi}^a}{\sqrt{\tau \, e^\varphi}} \,,
\end{align}
\begin{align}
  &m^a=\frac{1}{\sqrt{2}} \left( e_{(2)}^a + i e_{(3)}^a \right) \,,\non\\
  &\bar{m}^a=\frac{1}{\sqrt{2}} \left( e_{(2)}^a - i e_{(3)}^a \right) \,,
\end{align}
where~$e_{(2)}^a$ and~$e_{(3)}^a$ are real unit vectors orthogonal
to~$l^a+n^a$ and~$l^a-n^a$. From the Weyl tensor~$C_{abcd}$, one may
construct the Weyl scalars~$\Psi_4$, $\Psi_3$, $\Psi_2$, $\Psi_1$
and~$\Psi_0$, which are defined as follows:
\begin{align}
  &\Psi_4=C_{abcd} n^a \bar{m}^b n^c \bar{m}^d \,,\non\\
  &\Psi_3=C_{abcd} l^a n^b \bar{m}^c n^d \,,\non\\
  &\Psi_2=C_{abcd} l^a m^b \bar{m}^c n^d \,,\non\\
  &\Psi_1=C_{abcd} l^a n^b l^c m^d \,,\non\\
  &\Psi_0=C_{abcd} l^a m^b l^c m^d \,.
\end{align}
We say that the Weyl scalars~$\Psi_N$ satisfy the peeling property if,
for some appropriately defined radial parameter~$r$, the asymptotic
behavior of the Weyl scalars has the form~\cite{NewPen62}:
\begin{align}\label{Peeling}
  &\Psi_N \sim \frac{1}{r^{5-N}} \,.
\end{align}
It is worth mentioning that in the classical peeling result
in~\cite{NewPen62} the coordinates used are~$(u,r,\theta^A)$ where~$u$
satisfies the eikonal equation and~$l^a=g^{ab}\nabla_bu$ are tangent
to a family of null geodesics. Additionally, the freedom in rotating
the~$m$ and~$n$ legs of the tetrad is fixed by demanding that they are
parallel propagated along~$l^a = (\frac{\partial}{\partial r})^a$
where~$r$ is an affine parameter along the null
geodesics. Nonetheless, in other classical studies such as
in~\cite{Sac62} the areal radial coordinate (also called luminosity
parameter) $\tilde{r}$ has been used to perform similar asymptotic
expansions of the curvature tensors and the metric.  The relation
between Newman-Penrose's~$r$ and Bondi-Sach's~$\tilde{r}$ has been
derived in \cite{Val99} ---see also~\cite{MadJef16} for a discussion
of the Bondi-Sachs set up. However, in our case we will not use~$r$
nor~$\tilde{r}$ but work with the radial-Shell coordinate~$R$ as
defined in section~\ref{setup}. This is a subtle point since, in order
to fully compare an asymptotic expansion for the Weyl tensor to that
of the classical Peeling theorem of~\cite{NewPen62} one needs not only
to find the transformation between the coordinates but also between
the frames.  The precise relation between our set up (coordinates and
frame) and that of~\cite{NewPen62} will be analyzed elsewhere.  We
compute the Weyl scalars~$\Psi_N$ using the polyhomogeneous expansions
for each of the metric functions in the Cartesian harmonic gauge,
including terms up to order~$R^{-3}$ and logarithmic terms indicated
by~\eqref{FunctionalFormCart1},~\eqref{FunctionalFormCart2}
and~\eqref{FunctionalFormCart3}. We find that~$\Psi_4$ and~$\Psi_3$
satisfy the peeling property (Here, the parenthetical
subscripts~$(n,k)$ denoting the order of coefficients in the
polyhomogeneous expansion):
\begin{align}
  &\Psi_4 = -\frac{\left( \ddot{h}_{+(1,0)} + i \ddot{h}_{\times(1,0)} \right)}{R}
  + o(R^{-1}) \,,\non\\
  &\Psi_3 = \frac{1}{4 R^2}
  \biggl[
    - \partial_\theta \dot{h}_{+(1,0)} - i \partial_\theta  \dot{h}_{\times(1,0)}
    + \partial_\theta \dot{\varepsilon}_{(1,0)}
    - \partial_\theta \dot{\varphi}_{(1,0)}\non\\
    & \qquad
    -\ddot{C}^+_{\theta(1,0)} - i \csc(\theta)
    \biggl( \partial_\phi \dot{h}_{+(1,0)}
    + i \partial_\phi \dot{h}_{\times(1,0)}
    + \partial_\phi \dot{\varepsilon}_{(1,0)}\non\\ & \qquad
    - \partial_\phi \dot{\varphi}_{(1,0)} - \ddot{C}^+_{\phi(1,0)} \biggr)
    - 2 \cot(\theta) \left( \dot{h}_{+(1,0)} + i \dot{h}_{\times(1,0)} \right)
  \biggr]
   \non\\ & \qquad + o(R^{-2}) \,.
\end{align}
To compute~$\Psi_2$, it is convenient to introduce the following
contractions of the Ricci tensor, which vanish on vacuum solutions:
\begin{align}
  &\mathcal{H} = R_{ab} (l^a + n^a) (l^b + n^b) \,,
\end{align}
\begin{align}
  &\mathcal{M} = R_{ab} (l^a + n^a) (l^b - n^b) \,,
\end{align}
Recall that~$R_{ab}=\mathcal{R}_{ab}+\nabla_{a}Z_{b} - W_{ab}$, and if
the reduced vacuum EFE are imposed ---via the wave equations of
section~\ref{EFE}--- one has that~$\mathcal{R}_{ab}=0$. Thus,
requiring~$\mathcal{H}=0$ and~$\mathcal{M}=0$ is guaranteed by
assuming that the~$Z_a$ constraints are satisfied. Due to the
construction for the propagation of constraints, roughly speaking one
can view~$\mathcal{H}$ and~$\mathcal{M}$ as a subset of the
Hamiltonian and momentum constraints. We do not impose that any of the
constraints vanish but, inline with the free-evolution philosophy,
given a solution to the reduced field equations, we are free to add
and subtract any amount of those constraints to the equations when
calculating quantities after the fact. Upon adding the following
combination of~$\mathcal{H}$ and~$\mathcal{M}$, we obtain an
expression of the form
\begin{align}
  &\Psi^\prime_2 = \Psi_2
  - \frac{\mathcal{H}+\mathcal{M}}{12}  \non\\
  & \qquad
  = \frac{\Psi_{2(3,0)}}{12 R^3}
  + \frac{\log R}{4 R^3} \Psi_{2(3,1)} + o(R^{-3})\,,
\end{align}
where:
\begin{align}
  \Psi_{2(3,0)}
  =& ~-3 \mathcal{C}^R_{-(1,0)} + 2 \mathcal{C}^R_{-(1,1)}
  + 3 \mathcal{C}^R_{+(1,0)} + 2 h_{+(1,0)} - 6 \dot{\varepsilon}_{(2,0)}
  \non\\
  & + 4 \dot{\varepsilon}_{(2,1)}
  - \partial_\theta^2 \mathcal{C}^R_{+(1,0)}
  - \partial_\theta^2 h_{+(1,0)} + \partial_\theta^2 \varepsilon_{(1,0)}
  \non\\
  &  - 2 \partial_\theta^2 \varphi_{(1,0)} + 6 \dot{\mathcal{C}}^R_{+(2,0)}
  + 2\cot(\theta) \csc(\theta) \partial_\phi {h}_{\times(1,0)}
  \non\\
  & - 3 \left(h_{+(1,0)} - i h_{\times(1,0)}\right)
  \left(\dot{h}_{+(1,0)} + i \dot{h}_{\times(1,0)}\right)
  \non\\
  & - \cot(\theta)
  \biggl[
    \partial_\theta \mathcal{C}^R_{+(1,0)} + 3 \partial_\theta h_{+(1,0)}
    - \partial_\theta \varepsilon_{(1,0)}
    \non\\
    &
    + 2 \partial_\theta \varphi_{(1,0)}
    \biggr]
  + \csc^2(\theta)
  \biggl[
    \partial_\phi^2 h_{+(1,0)} -  \partial_\phi^2 \mathcal{C}^R_{+(1,0)}
    \non\\
    & + \partial_\phi^2 \varepsilon_{(1,0)} - 2 \partial_\phi^2 \varphi_{(1,0)}
    + 2 \sin(\theta) \partial_\theta \partial_\phi h_{\times(1,0)}
    \biggr],
  \non\\
  \Psi_{2(3,1)}
  =&~
  - \mathcal{C}^R_{-(1,1)} + 2 \dot{\mathcal{C}}^R_{+(2,1)}
  - 2 \dot{\varepsilon}_{(2,1)}\,.
\end{align}
Since the coefficient~$\mathcal{C}^R_{-(1,1)}$ appears
in~$\Psi_{2(3,1)}$, the Weyl scalar will fall off as~$\log R/R^3$. It
follows that for solutions of the asymptotic system in Cartesian
harmonic coordinates, and with this particular constraint
addition~\eqref{ConstAdd}, the Weyl scalar~$\Psi_{2}$ fails to
satisfy~\eqref{Peeling}. We can then conclude that the vacuum EFE in
GHG do not, in general, satisfy the peeling property (even when
working within our restricted class of initial data). However we shall
see in the next section that it is possible to pick the gauge choice
and constraint addition carefully in order to recover this property.

\section{Recovering peeling}\label{recoverpeeling}

As previously said, the asymptotic behavior of solutions of GR near
null infinity is affected by a special interplay between the gauge and
the constraint addition we choose to work with. We have seen in the
previous section that our first choice results in a violation of the
peeling property of the Weyl scalars, even within our special class of
initial data. In this section we will see however that it is possible,
with the same type of initial data, to find a gauge and constraint
addition in which peeling is manifest. This is done by finding that
there is at least one choice which makes 2 of the metric functions
behave as goods and 8 as uglies. Then, by Remark~\ref{remark1}, one
can make the first logs appear only at arbitrarily high order, so that
the Weyl scalars cannot possibly contain~$\log R$ to leading order.

\paragraph*{Gauge choice:} We choose a gauge that is Cartesian harmonic
to leading order, and add a higher order correction~$\check{F}^a$,
\begin{align}\label{F=Fr+Fc}
F^a = \ring{F}^a+\check{F}^a\,.
\end{align}
Before choosing~$\check{F}^a$ we will compute the first order asymptotic
system to find the most convenient choice for our purposes.

\paragraph*{Constraint addition:} For the same reason that led us to
keeping $\check{F}^a$ free, we want to keep some freedom in the
constraint addition to choose {\it a posteriori}. We parameterize that
freedom by the four natural numbers~$p_1$, $p_2$ and~$p_A$, and we
make the following constraint addition,
\begin{align}\label{ConstAdd2}
  &W_{\psi\psi} = -\frac{1}{2}(Z^\sigma)^2+\frac{p_1}{R}Z^\sigma\,,\non\\
  &\slashed{W} = -\frac{p_2}{R}Z^{\sigmab}\,,\non\\
  &W_{\psi A} = \frac{3-p_A}{R}Z^A\,.
\end{align}

\paragraph*{GHG in good-bad-ugly form:} Our choice of gauge,
together with constraint addition~\eqref{ConstAdd2}, gives the
following system,
\begin{align}
  &  \mathring{\square} \varphi
  = \nabla_T\check{F}^\sigma+\mathcal{N}_\varphi\,, \non \\
  & \mathring{\square} \mathcal{C}_{+}^R
  = \frac{2p_1}{R}\nabla_T\mathcal{C}_+^R
  + \mathcal{N}_{\mathcal{C}_{+}^R} \non\,,\\
  &\mathring{\square} \mathcal{C}_{-}^R
  = - \frac{1}{2}(\nabla_T h_+)^2
  -\frac{1}{2}(\nabla_T h_\times)^2
  -2\nabla_T\check{F}^{\sigmab}
  +\mathcal{N}_{\mathcal{C}_{-}^R}\,,\non \\
  &\mathring{\square} \hat{\mathcal{C}}_A^+  =
  \frac{2p_A}{R}\nabla_T\hat{\mathcal{C}}_A^+
  +\mathcal{N}_{\hat{\mathcal{C}}_A^+} \,,\non\\
  &\mathring{\square} \hat{\mathcal{C}}_A^-
  = \frac{4}{R}\nabla_T\hat{\mathcal{C}}_A^-
  -2R\nabla_T\check{F}^A +\mathcal{N}_{\hat{\mathcal{C}}_A^-}\,, \non\\
  &\mathring{\square} \epsilon
  = \frac{2p_2}{R}\nabla_T\epsilon +\mathcal{N}_{\epsilon}\,,\non\\
  &\mathring{\square} h_+ = \mathcal{N}_{h_+}\,,\non \\
  &\mathring{\square} h_\times = \mathcal{N}_{h_\times}\,.
\end{align}
A close look shows that~$h_+$ and~$h_\times$ satisfy good equations
regardless of the freedom we still have. So we are left with~$8$
equations and~$8$ degrees of freedom given by the scalar
functions~$\check{F}^a$ and the choice of constraint addition,
paramaterized here by the natural numbers~$p_1$, $p_2$ and~$p_A$, all
of which may influence the asymptotic solution space. We want to
recover the peeling property of the Weyl scalars, so our criteria in
choosing our free functions will be to have as few logs in the first
few orders as possible. In fact, there is a particular choice which
makes the expansion of the metric functions log-free to arbitrarily
high order. As was stated in Remark~\ref{remark1}, the polyhomogeneous
expansion of a field which satisfies an ugly equation of the
type~\eqref{gbugeneral}, provided that no logs are inherited through
coupling with other equations, will have no logs up to order
$p$. Therefore our strategy here is to set~$p_1=p_2=p_A=p$ and
choose~$p$ large enough to get rid of all logs in the ugly fields that
may contribute to the leading decay in the Weyl scalars. However we
must guarantee that no logs are inherited through coupling, and we
will do that by carefully choosing~$\check{F}^a$ in order to
turn~$\varphi$, $\mathcal{C}_-^R$ and~$\hat{\mathcal{C}}_A^-$ into
uglies with no logs to order~$p$. Therefore, we want~$\check{F}^a$ to
satisfy the following conditions,
\begin{align}\label{FcConditions}
  &\nabla_T\check{F}^\sigma \simeq
  \frac{2p}{R}\nabla_T\varphi\,,\non\\
  &\nabla_T\check{F}^{\sigmab}
  \simeq- (\nabla_T h_+)^2-(\nabla_T h_\times)^2
  - \frac{p}{R}\nabla_T\mathcal{C}_-^R\,,\non\\
  &\nabla_T\check{F}^A \simeq
  -\frac{p-2}{R^2}\nabla_T\hat{\mathcal{C}}_A^-\,,
\end{align}
so that the system turns into,
\begin{align}
  &  \mathring{\square} \varphi =
  \frac{2p}{R}\nabla_T\varphi+\mathcal{N}_\varphi\,, \non\\
  & \mathring{\square} \mathcal{C}_{+}^R  =
  \frac{2p}{R}\nabla_T\mathcal{C}_+^R+ \mathcal{N}_{\mathcal{C}_{+}^R} \non\,,\\
  &\mathring{\square} \mathcal{C}_{-}^R  =
  \frac{2p}{R}\nabla_T\mathcal{C}_-^R+ \mathcal{N}_{\mathcal{C}_{-}^R}\,,\non \\
  &\mathring{\square} \hat{\mathcal{C}}_A^+  =
  \frac{2p}{R}\nabla_T\hat{\mathcal{C}}_A^+
  +\mathcal{N}_{\hat{\mathcal{C}}_A^+} \,,\non\\
  &\mathring{\square} \hat{\mathcal{C}}_A^-  =
  \frac{2p}{R}\nabla_T\hat{\mathcal{C}}_A^-
  +\mathcal{N}_{\hat{\mathcal{C}}_A^-}\,, \non\\
  &\mathring{\square} \epsilon  =
  \frac{2p}{R}\nabla_T\epsilon +\mathcal{N}_{\epsilon}\,,\non\\
  &\mathring{\square} h_+ = \mathcal{N}_{h_+}\,,\non \\
  &\mathring{\square} h_\times = \mathcal{N}_{h_\times}\,.
\end{align}
With this we would be able to write the~$10$ metric functions as~$2$
goods and~$8$ uglies. Moreover, $p$ has not yet been chosen so by
Remark~\ref{remark1}, we are allowed to have logs only from an
arbitrarily high order onward. Note that~\eqref{FcConditions} is only
an asymptotic condition, meaning that we might be interfering with
terms that contribute to sub-leading orders. However, our goal here is
to guarantee that our equations are only either good or ugly, and that
is determined solely by the first order asymptotic system. In the
following we explain how~\eqref{FcConditions} can be achieved.

\paragraph*{Making sure hyperbolicity is maintained:} There is a
subtlety in guaranteeing that the gauge source functions
satisfy~\eqref{FcConditions}. In order to write $\check{F}^a$ in terms
of metric functions, we can simply integrate the first and last
equations to get,
\begin{align}
  \check{F}^\sigma = \frac{2p}{R}(e^\varphi-1)\,,\quad
  \check{F}^A=\frac{2-p}{R^2}\hat{\mathcal{C}}_A^-\,.
\end{align}
This simple-minded approach to the second equation fails, as it does
not integrate cleanly, and instead yields an expression
for~$\check{F}^{\sigmab}$ involving an integral of derivatives
of~$h_+$ and~$h_\times$. To overcome the issue we take inspiration
from the gauge driver approach of~\cite{LinMatRin07}. We begin by
separating~$\check{F}^{\sigmab}$ into two pieces, an explicit one and
a gauge driver~$\check{F}^{\sigmab}_1$, as follows:
\begin{flalign}
 \check{F}^{\sigmab} = \frac{1}{R}\check{F}^{\sigmab}_1 -
 \frac{p}{R}(1+\mathcal{C}_-^R) \,,
\end{flalign}
so that, according to~\eqref{FcConditions}, $\check{F}^{\sigmab}_1$
must now satisfy,
\begin{align}\label{Fc1}
  \frac{1}{R}\nabla_T\check{F}^{\sigmab}_1 \simeq
  - (\nabla_T h_+)^2-(\nabla_T h_\times)^2 := \frac{1}{R^2}H \,.
\end{align}
Since~$h_+$ and~$h_\times$ are goods, we can write,
\begin{align}\label{Fc12}
  H \simeq H_{1,0}(\psi^*) \,.
\end{align}
Our strategy consists of insisting that~$\check{F}^{\sigmab}_1$
satisfies a wave equation whose solution has the property
that~\eqref{Fc1}. Then, second derivatives of~$\check{F}^{\sigmab}_1$
in wave equations for metric functions can be treated as a variable to
be evolved instead of as functions of the metric themselves. We impose
the following gauge-driving evolution equation
\begin{align}\label{FcWaveEq}
  \mathring{\square}\check{F}^{\sigmab}_1 =
  \frac{2q}{R}\left(\nabla_T \check{F}^{\sigmab}_1
  - \frac{1}{R}H\right) + \mathcal{N}_{\check{F}^{\sigmab}_1}\,,
\end{align}
where~$q$ is a natural number deliberately left free for the
moment. The reason for the specific form of this equation will become
clear later in this section, but {\it morally} one can already think
of the first term on the RHS as suppressing the part of the radiation
field of~$\check{F}^{\sigmab}_1$ that would naturally appear from
initial data, and the second term forcing the radiation field equal to
a desired target. We included a combination of stratified null
forms~$\mathcal{N}_{\check{F}^{\sigmab}_1}$ because, as they do not
contribute to leading order, they can be chosen in order to make the
wave equation as simple as possible. We shall make that choice
later. Rescaling our problematic function
as~$f_1:=R\check{F}^{\sigmab}_1$ then gives,
\begin{align}\label{FcWaveEqResc}
  -\frac{2}{R}\nabla_\psi\nabla_T
  f_1=\frac{2q}{R^2}\left(\nabla_T f_1 - H\right) +
  \mathcal{N}'_{\check{F}^{\sigmab}_1}\,,
\end{align}
where~$\mathcal{N}'_{\check{F}^{\sigmab}_1}$ is
just~$\mathcal{N}_{\check{F}^{\sigmab}_1}$ minus the stratified null
forms coming from the LHS. We now add~$2R^{-1}\nabla_\psi H$ on both
sides,
\begin{align}
  -\frac{2}{R}\nabla_\psi\left(\nabla_Tf_1-H\right)=&
  \frac{2q}{R^2}\left(\nabla_T
  f_1 - H\right) \\&+\frac{2}{R}\nabla_\psi H+
  \mathcal{N}'_{\check{F}^{\sigmab}_1}\,.\non
\end{align}
As we left~$\mathcal{N}_{\check{F}^{\sigmab}_1}$ free, we can still
choose it so that~$\frac{2}{R}\nabla_\psi H = -
\mathcal{N}'_{\check{F}^{\sigmab}_1}$, since~$\nabla_\psi H$ is
necessarily a stratified null form~\eqref{Fc12}. This is not an
essential step, but it makes the resulting equation more
tractable. Therefore we get the simple equation,
\begin{align}
  \nabla_\psi\left(\nabla_Tf_1-H\right)=-\frac{q}{R}\left(\nabla_T
  f_1 - H\right)\,,
\end{align}
that we can integrate to find,
\begin{align}\label{delTf1-H}
  \nabla_Tf_1-H=\frac{1}{R^q}\alpha(\psi^*)\,.
\end{align}
From~\eqref{delTf1-H} we see that as long as $q$ is a natural number,
the condition~\eqref{Fc1} is satisfied, and we can
turn~$\mathcal{C}_-^R$ into an ugly while ensuring that hyperbolicity
is maintained. However, by introducing an eleventh non-linear wave
equation, we may be introducing logs in the system, which will then
appear in the EFE through coupling. In fact, the wave equation we
forced~$\check{F}^{\sigmab}_1$ to satisfy, namely~\eqref{FcWaveEq}
does not fit directly into our conception of good, bad or ugly. For
that reason, we have to show that~$\check{F}^{\sigmab}_1$ itself
admits a polyhomogeneous expansion in which logs only appear at
arbitrarily high order. We rescale~$f_1:=R\check{F}^{\sigmab}_1$ and
assume that~$f_1$ satisfies the conditions~\eqref{weaknull1}, so we
can write,
\begin{align}\label{FcWaveEqResc2}
  \nabla_\psi\left(R^q\nabla_T f_1\right)\simeq qR^{q-1}H_{1,0}\,,
\end{align}
Integrating we get,
\begin{align}
	f_1\simeq F_{1,0}(\psi^*)\,,
\end{align}
where~$F_{1,0}(\psi^*)$ is constant along~$\psi^a$. Now we assume that,
\begin{align}\label{Fn-1}
  \check{F}^{\sigmab}_1 = \frac{F_{1,0}(\psi^*)}{R} +
  \sum_{m=2}^{n-1}\sum_{k=0}^{N_{n-1}^F} \frac{(\log R)^k
    F_{m,k}(\psi^*)}{R^m}+\frac{\mathcal{F}_n}{R^n}\,,
\end{align}
where~$\mathcal{F}_n=o^+(R)$, and must demonstrate,
\begin{align}\label{Fn}
  \check{F}^{\sigmab}_1 = \frac{F_{1,0}(\psi^*)}{R} +
  \sum_{m=2}^{n}\sum_{k=0}^{N_n^F} \frac{(\log R)^k
    F_{m,k}(\psi^*)}{R^m}+\frac{\mathcal{F}_{n+1}}{R^{n+1}}\,.
\end{align}
Plugging~\eqref{Fn-1} into~\eqref{FcWaveEq} and collecting terms
proportional to~$R^{-n-1}$ we get,
\begin{align}\label{Fmot3}
\nabla_\psi(R^{q+1-n}\nabla_T\mathcal{F}_n) \simeq
R^{q-n}\Omega^F_{n-1}\,,
\end{align}
where,
\begin{align}\label{expH2}
\Omega^{F}_{n-1} = \sum_{i=0}^{N_n^{\Omega^F}}(\log
R)^i\Omega^{F}_{n-1,i}(\psi^*)\,.
\end{align}
Integrating~\eqref{Fmot3} we find that~\eqref{Fn} and so, by
induction, $\check{F}^{\sigmab}_1$ admits a polyhomogeneous expansion
near null infinity of the type,
\begin{align}
  \check{F}^{\sigmab}_1 = \frac{F_{1,0}(\psi^*)}{R} +
  \sum_{m=2}^{\infty}\sum_{k=0}^{N_n^F} \frac{(\log
    R)^kF_{m,k}(\psi^*)}{R^m}\,.
\end{align}
To this point we have allowed~$\check{F}^{\sigmab}_1$ to have any
power of~$\log R$ in terms beyond first order. However, an argument
completely analogous to Remark \ref{remark1} can be made here in order
to show that there can be no logs up to order~$n=q$ as long as no logs
are inherited through coupling. With this in mind, we choose~$q=p$ and
notice that we thus avoid the danger of introducing logs in the first
orders of the asymptotic expansions. By insisting
that~$\check{F}^{\sigmab}_1$ satisfies the wave
equation~\eqref{FcWaveEq}, we have prevented our special choice of
gauge from interfering with hyperbolicity.  Finally, we have
effectively shown that a choice of gauge and constraint addition
allows us to have~$2$ goods and~$8$ uglies with no logs up to
arbitrarily high order. We conclude this section with a paragraph on
how this result recovers the peeling property.

\paragraph*{Recovering peeling:} As we saw in section~\ref{NoPeeling},
the peeling property is satisfied iff,
\begin{align}
	&\Psi_N \sim \frac{1}{R^{5-N}} \,,
\end{align}
where~$N\in \{0,1,2,3,4\}$. We want to use our freedom in the choice
of the natural number~$p$ to make sure that logs only appear in the
polyhomogeneous expansion at an order that is high enough that they
cannot possibly appear in the leading orders of the Weyl
scalars. These are components of the Weyl tensor and hence they are
non-linear combinations of the metric and its first and second
derivatives. Since no derivatives can ever decrease the order of the
given term in $R^{-1}$, we need only know the functional form of the
metric components to order $R^{-5}$ in order to determine whether the
peeling property is satisfied. We need to make sure that the metric
components do not contain any logs to order $n=5$, or rather,
\begin{align}
  g_{ab} = \sum_{n=1}^{5}\frac{G_{ab,n}(\psi^*)}{R^n}
  + \frac{\mathcal{G}_6}{R^6}\,,
\end{align}
where~$\mathcal{G}_6=o^+(R)$. Now in order to choose~$p$, we must take
into account that certain metric components contain factors of~$R$
and~$R^2$ attached to metric functions that are themselves good, bad
or ugly, as can be seen from~\eqref{eq:qconformaltransformation}
and~\eqref{asympflatness}. That said, without actually computing the
dependence of the Weyl scalar on the metric functions, we can
guarantee that they will not contain any logs to leading order if we
require that the metric functions do not contain any logs to order
$n=7$. This is achieved by choosing~$p\geq7$, in accordance with
Remark~\ref{remark1}.  So, by carefully choosing the gauge and
constraint addition, we have prevented the existence of any terms
proportional to~$\log R$ in the polyhomogeneous expansions of all the
metric functions up to order~$n=7$, thereby recovering, within our
class of initial data, the peeling property of the gravitational field
in the EFE-GHG setup.

\section{Recovering logs through a coordinate change}\label{logsback}

In the previous sections we saw that the Cartesian harmonic gauge,
together with the constraint addition that turns 4 of the metric
functions into uglies, generates terms proportional to~$\log R$ at
first order in~$\cmr$ and at second order in some of the other
functions. These logs in turn, give rise to a violation of
peeling. Then it was shown that by changing the gauge to a very
specific one and modifying the constraint additions slightly to
contain the natural number~$p$, we can delay the appearance of those
logs to an arbitrarily high order. This ensures that peeling is
recovered if~$p$ is high enough. Here we will not be concerned with
peeling, but with how these logs are generated. So for simplicity we
fix~$p=1$ so that the constraint additions in either case are exactly
the same, and focus on the log that appears at leading order, i.e. the
one in $\cmr$. In that case, the EFE only differ by the choice of
gauge, so we should, in principle, be able to generate and get rid of
the logs just by a particular coordinate change. To illustrate this,
we begin with a coordinate system~$\{T,R,\theta^A\}$ which we assume
generates no logs to leading order.  Then we change one coordinate,
say~$R$, to something different, $\{T,\rho,\theta^A\}$, and check that
the leading log in $\cmr$ is recovered. The relation between the
new~$\mathcal{C}^\rho_-$ and the old~$\cmr$ is given by,
\begin{align}\label{cmrcmrho}
  \mathcal{C}^\rho_- = \cmr J^\rho{}_R + J^\rho{}_T = \nabla_{\psib}\rho\,,
\end{align}
which may be obtained by comparing expressions for the covector
basis~$\sigma_a$, $\ul{\sigma}_a$ between the two sets of coordinates.
The wave equation that~$\rho$ satisfies is,
\begin{align}\label{waveeqrho}
	\square \rho - F^\rho =0\,,
\end{align}
and the most straightforward way to find the asymptotics of $\rho$
through~\eqref{waveeqrho} is by taking advantage of the analysis we
already have of the good, bad and ugly equations. However, the
variable~$\rho$ is expected to grow like~$R$, so it is not exactly a
variable like the~$g$, $b$ or~$u$. So we choose to work with~$\rho/R$
instead and change~\eqref{waveeqrho} accordingly,
\begin{align}
  R\square \frac{\rho}{R} = -2g^{aR}\nabla_a \frac{\rho}{R}
  + \frac{\rho}{R}\Gb^R + F^\rho\,.
\end{align}
This can be written as,
\begin{align}
  \mathring{\square} \frac{\rho}{R}
  = \frac{2}{R}\nabla_T\frac{\rho}{R}+\mathcal{N}_{\frac{\rho}{R}}\,,
\end{align}
which implies that~$\frac{\rho}{R}$ is an ugly and admits a
polyhomogeneous expansion of the form,
\begin{align}
  \frac{\rho}{R} = 1 + \frac{m(\theta^A)}{R} + o^+(R^{-1})\,,
\end{align}
which, for convenience, we will write as,
\begin{align}
  \rho = R + m(\theta^A) + \frac{\Phi}{R}\,,
\end{align}
by defining the quantity~$\Phi$ accordingly. If we plug this
in~\eqref{waveeqrho} and use constraint addition, we get,
\begin{align}
  \square\frac{\Phi}{R} = F^\rho - F^R - \cancel{\Delta}m=o^+(R^{-1})\,,
\end{align}
which implies,
\begin{align}
  -\frac{1}{R}\nabla_\psi\nabla_{\psib}\Phi
  = F^\rho - F^R - \cancel{\Delta}m +o^+(R^{-2})\,.
\end{align}
Upon integration along integral curves of~$\psi^a$, the only way
for~$\nabla_{\psib}\Phi$ not to have a log is if,
\begin{align}
  F^\rho - F^R - \cancel{\Delta}m = o^+(R^{-2})\,,
\end{align}
which, in general, is not true. In fact, one of the second order terms
in~$F^R$ is,
\begin{align}\label{loggenterm}
  \frac{1}{R}\check{F}^{\sigmab}_1 \simeq
  - \int_{T_0}^T\left[(\nabla_T h_+)^2+(\nabla_T h_\times)^2\right]dT'\,,
\end{align}
precisely the term that gives rise to the log at first order
in~$R^{-1}$ in the polyhomogeneous expansion of~$\cmr$ in the
Cartesian harmonic gauge. So we see that if~$F^\rho -
\cancel{\Delta}m$, both determined by our choice of coordinates, do
not contain a term symmetric to~\eqref{loggenterm}, a log is
unavoidable. Therefore, in general,
\begin{align}
  \rho = R + m(\theta^A)
  + \frac{\Phi_{1,0}(\psi^*)
    +\Phi_{1,1}(\psi^*)\log R}{R} + o^+(R^{-1})\,,\non
\end{align}
which, together with~\eqref{cmrcmrho} gives,
\begin{align}
  \mathcal{C}^\rho_- = \cmr + \frac{\nabla_{\psib}\Phi_{1,0}
    +\nabla_{\psib}\Phi_{1,1}\log R}{R} + o^+(R^{-1})\,,
\end{align}
thus recovering a~$\log R$ to first order in~$R^{-1}$ as we wanted. 

\section{Conclusions}\label{conclusions}

Consider initial data that is, in some sense, asymptotically flat and
that generates a spacetime with a piece of future null
infinity. Introduce coordinates and a null-frame appropriately adapted
to asymptotic flatness in a neighborhood of null infinity. If the Weyl
scalars decay like~$\Psi_N=O(r^{N-5})$, with~$r$ a radial coordinate
naturally constructed from the adapted coordinates, then the Weyl
tensor is said to peel. Whether or not the Weyl tensor peels is
intimately tied both to the specific notion of asymptotic flatness
imposed on initial data, and to the coordinates and frame
employed. Described as such, the notion of peeling appears rather
pedestrian. It could be that even within a strict class of initial
data one could conclude that the Weyl tensor both peels and does not
peel, depending on the choice of coordinates and frame. A refined
formulation, for which we have much sympathy, would be to say that the
Weyl tensor peels if any set of coordinates and frame can be found in
which the aforementioned decay is manifest. In practice it is often
the more pedestrian notion of peeling that is of concern, since that
is directly related to smoothness at null-infinity within the gauge
employed. Besides which, demonstrating that no coordinates exist in
which peeling holds is a much less concrete task than examining
specific cases of interest.

A fundamental ingredient in the small-data global existence result in
harmonic gauge~\cite{LinRod04} was to understand and control terms in
the metric components of the asymptotic form~$\log R / R$ near null
infinity. Since such terms decay more slowly than solutions to the
wave equation and appear in the presence of any gravitational wave
content, the questions arise: could the peeling property generically
be satisfied when using harmonic gauge, even with a very strict notion
of asymptotic flatness? Assuming that this is not the case, could
peeling be recovered for the same type of initial data by using a
suitable generalized harmonic gauge?

The main achievement of this work was to address these questions by
exploiting the formal asymptotic expansions of~\cite{DuaFenGasHil21}.
We found that there are choices of gauge and constraint addition which
give rise to a violation of the peeling property. For this it was
necessary to generalize the last theorem of~\cite{DuaFenGasHil21} to
include systems of any number of good, bad and ugly equations with
stratified null forms, as well as to generalize the notion of
`ugly'. This was to establish that such a system admits formal
polyhomogeneous asymptotic solutions near null infinity for a broad
class of initial data. Since in general the metric
components~$g_{\ul{\mu\nu}}$ with respect to the GHG coordinate
basis~$\partial_{\ul{\mu}}$ does not fall into our good-bad-ugly
categorization, these methods cannot be directly and cleanly applied
to the standard formulation of the EFE in GHG (in
which~$g_{\ul{\mu\nu}}$ are the variables). However, exploiting that
this categorization and the first order system discussed
in~\cite{LinRod03} and~\cite{GasHil18} are based on the privileged
role played by null directions, we have introduced a reformulation of
the EFE in GHG where the variables are written in terms of the
coordinate lightspeeds (encoding the non-trivial components of the
outgoing and incoming null vectors). One point to notice here is that
in~\cite{LinRod03} and~\cite{GasHil18} the null directions are those
of the Minkowski spacetime while in the formulation discussed here,
the null directions are the (true) null directions
in~$(\mathcal{M},g)$. This special set of variables has the property
that, the evolution equations in the EFE in GHG, upon a choice of
gauge and constraint addition, have the structure of a good-bad-ugly
system.

This was carefully seen in section~\ref{AsCartHarm} by considering the
Cartesian harmonic gauge and then adding constraints in order to
maximize the number of uglies. Analyzing the asymptotic system via
computer algebra, it was possible to use the theorem shown in
section~\ref{GBUwSNF} to find the maximum power of~$\log R$ that each
variable may have at each order up to 3rd in~$R^{-1}$. Computing the
leading order terms in the Weyl scalars, we were able to assert that,
at least within the formal expansions framework
of~\cite{DuaFenGasHil21} the peeling property is not, in general,
satisfied by solutions in this gauge, as these contain leading terms
proportional to~$\log R$, instead of only negative integer powers
of~$R$.

Hence, physically relevant spacetimes evolved (say numerically using
the EFE in GHG), will not, in general, possess a smooth null infinity
in that gauge. We saw, however that it is possible to choose the gauge
and constraint addition carefully to guarantee that terms proportional
to~$\log R$ do not appear in the few first orders of the metric
components, hence recovering peeling within the same class of initial
data. Interestingly, in contrast to the order-by-order approach used
for harmonic gauge in the post-Minkowskian setting~\cite{Bla86} the
generalized harmonic strategy we apply suppresses~$\log R$ terms down
to a finite order all at once, and may therefore be of use in that
context.

It is furthermore interesting that the violation of peeling identified
in this paper occurs for~$\Psi_2$ and is of the form~$\log R
/R^3$. This formally coincides with the violation of peeling reported
in~\cite{GasVal18}. But unlike the present work, the result
of~\cite{GasVal18} was given in the Newman-Penrose
gauge~\cite{NewPen62}. Establishing a relation between the gauges used
in this paper and other gauges such as the Newman-Penrose gauge or the
F-gauge of~\cite{fri98a} is of deep interest but requires a delicate
analysis that will be done elsewhere. In the framework of the formal
expansions of solutions to the CEFEs discussed in~\cite{GasVal18}, see
also~\cite{fri98a}, the violation of peeling can be related to a
regularity condition at the level of the initial data in terms of
derivatives of the Bach tensor of the initial conformal metric
evaluated at spatial infinity. The main focus of this paper has been
on the evolution equations and the choice of gauge. However, the
relation between the class of initial data, solutions to the
Hamiltonian and momentum constraints, and the terms appearing in our
asymptotic expansions, and ultimately in those responsible for the
violations of peeling, is of particular interest and deserves a full
study in itself, so is left for future work. It should be stressed
that the asymptotic expansions derived here as well as those reported
in~\cite{GasVal18} are formal and the relation between the expansions
and actual solutions is yet to be established. Nonetheless, the proof
of the global stability of the Minkowski spacetime of~\cite{ChrKla93}
predicts a decay for the Weyl scalars which, despite not being
polyhomogeneous, does not follow the fall-off predicted in the
classical peeling theorem~\cite{NewPen62}, and hence shows the failure
of peeling for generic initial data. The classical peeling result in
the setup of~\cite{ChrKla93} can be retrieved by restricting the
initial data as shown in~\cite{KlaNic03}. See also~\cite{HinVas20} for
a proof of the nonlinear stability of the Minkowski spacetime with
polyhomogeneous initial data leading to polyhomogeneous spacetime metric
developments. Perhaps one interpretation of the results here is that
the violation of peeling induced by the~$\log R/R$ terms in the metric
are, in a sense, {\it pure gauge}. The expectation is that within a
larger class of initial data, peeling will fail in the more refined
sense, that is, in such a way that it cannot be restored by a simple
change of gauge.

In this paper, the expected decay of the Weyl scalars according to the
classical peeling theorem was recovered in
section~\ref{recoverpeeling} by finding a particular choice that turns
our system of 10 non-linear wave equations into 2 goods and 8 uglies
with no logs up to arbitrarily high order. In the process, we have
built a method to find the functional form of polyhomogeneous
asymptotic solutions to the EFE in GHG near null infinity. In other
words, we now know the powers of~$\log R$ that may appear in the
metric at each order in~$R^{-1}$. This knowledge may allow us to
explicitly regularize formally singular terms appearing in a
hyperboloidal initial value formulation of the EFE in
GHG~\cite{Hil15,HilHarBug16,GasHil18,GauVanHil21} by a non-linear
change of variables like the one in~\cite{GasGauHil19}. We expect this
to be an important step towards a full regularization of the EFE in
GHG in the hyperboloidal setup, and that this might result. The full
details of this regularization will be presented elsewhere.\\

\acknowledgments

The Authors wish to thank Alex Va\~{n}\'o-Vi\~{n}uales for helpful
discussions. Many of our derivations were performed in
xAct~\cite{xAct_web_aastex} for Mathematica. The notebooks are
available at~\cite{DuaFenGas22_web}. MD acknowledges support from FCT
(Portugal) program PD/BD/135511/2018, DH acknowledges support from the
FCT (Portugal) IF Program IF/00577/2015, PTDC/MAT- APL/30043/2017. JF
acknowledges support from FCT (Portugal) programs
PTDC/MAT-APL/30043/2017, UIDB/00099/2020. EG gratefully acknowledges
support from the FCT (Portugal) 2020.03845.CEECIND and from the
European Union’s H2020 ERC Consolidator Grant “Matter and Strong-Field
Gravity: New Frontiers in Einstein’s Theory,” Grant Agreement
No. MaGRaTh-646597 and the PO FEDER-FSE Bourgogne 2014/2020
program/EIPHI Graduate School (contract ANR-17-EURE-0002) as part of
the ISA 2019 project.

\normalem
\bibliography{Peeling.bbl}{}

\end{document}